\documentclass[%
reprint,
nofootinbib,
 amsmath,amssymb,
 aps,
]{revtex4-2}

\usepackage{physics}
\usepackage{braket}
\usepackage{romanbar}
\usepackage{graphicx}
\usepackage[caption=false]{subfig}
\usepackage{dcolumn}
\usepackage{tikz}
\usepackage{bm}
\usepackage{float}
\usepackage{xcolor}
\usepackage[symbol,hang,flushmargin]{footmisc}
\usepackage{bbold} 
\begin{document}

\preprint{APS/123-QED}


\title{Photonic multipartite entanglement in discrete variables without arbitrary unitaries}

\author{Milica Bani\'c$^{1,*}$}
\footnotetext[1]{Milica.Banic@nrc-cnrc.gc.ca}
\author{J. E. Sipe$^2$}
\author{Marco Liscidini$^3$}
\affiliation{$^1$ National Research Council of Canada, 100 Sussex Drive, Ottawa, Ontario K1A 0R6, Canada\\$^2$ Department of Physics, University of Toronto, 60 St. George Street, Toronto, ON, M5S 1A7, Canada\\ $^3$ Department of Physics, University of Pavia, Via Bassi 6, 27100, Pavia, Italy}

\date{\today}

\begin{abstract}
{We present an approach for designing sources of postselected multipartite states based on photon-pair sources. Our approach can be applied to arbitrary target states in different encoding schemes and physical platforms. It also allows one to limit the types of components to be used in the device, such that lossy or difficult-to-implement optical elements can be avoided. As an example, we apply this strategy to design a passive integrated source of frequency-bin-encoded high-dimensional GHZ states with a 10 kHz on-chip generation rate for picojoule pump pulses.}
\end{abstract}

\maketitle


\section{Introduction}
\label{section:introduction}

Entangled states are necessary for test{s} of fundamental physics, and for many applications under the umbrella of ``quantum technologies" \cite{Pan2000_GHZ}. A simple example is the class of bipartite entangled states consisting of two particles entangled in a discrete degree of freedom. 
A much broader class of resources includes multipartite entangled states, which consist of three or more particles entangled in a manner that cannot be reproduced by ensembles of bipartite states. These multipartite states are important resources in practical applications including multipartite secret sharing, measurement-based quantum computation, and photonic quantum repeaters \cite{PhysRevA.59.1829_GHZ_secret_sharing,Azuma2015,PRXQuantum.4.040323,PhysRevA.68.022312}.
In many contexts, using entangled states of light in particular has a number of advantages; our focus here is on the generation of {photonic multipartite entangled states.}

While entangled pairs of photons can be generated directly by processes {such as}  
spontaneous four-wave mixing (SFWM) or spontaneous parametric down-conversion (SPDC), the generation of multipartite states is more challenging. 
Usually this is done by multiplexing many photon-pair sources, {employing} linear components {in the device,} and {implementing} postselection. {Such} 
devices require indistinguishability over the entire {structure,} 
and stability over the longest possible time scales. Integrated photonic circuits are a particularly promising platform for the implementation of these devices, due to their stability, scalability, tunability, and efficiency \cite{Bao2023,VLSI_doi:10.1126/science.aar7053}. 

%
%

There are many encoding schemes {from which} one can choose. 
Here we focus on {\textit{frequency bin encoding}}, a type of energy encoding scheme in which the logical states are sufficiently close 
in frequency that they can be modulated with commercial electro-optic modulators (EOMs) \cite{Lu:23}. This encoding scheme is robust, high-dimensional, and it is compatible with integrated photonic devices. Furthermore, a frequency bin encoding scheme is {\textit{scalable}} in the sense that the dimensionality of the photonic state can be increased without necessarily increasing the physical size of a device, unlike {approaches based on} 
path-encoding. 
Frequency bin encoding also has practical advantages in the context of networking and communications, since it is a resilient degree of freedom. 
For these reasons, frequency bin encoding in integrated structures seems to be a promising approach for the generation of entangled states in discrete variables. The generation of Bell states and higher-dimensional bipartite states has been demonstrated with high fidelity using integrated structures \cite{PhysRevLett.129.230505, Clementi2023, Borghi_PhysRevApplied.19.064026}, and the generation of multipartite states is {now being} 
explored \cite{banic_PhysRevA.109.013505}. 

Progress in this direction would benefit from a method for designing sources of multipartite states. One approach is to adapt approaches that have been established in other encoding schemes; for example, strategies for generating polarization-encoded GHZ and W states have been well-established for decades \cite{PhysRevLett.86.4435,W_PhysRevLett.92.077901}, and these have been ``translated" into other encoding schemes as well  \cite{Bergamasco_PRApp,Lo2023,Fang_PRL}. But the usefulness of this approach is limited: There are many states, particularly high-dimensional ones, for which generation schemes are not well-established. {And even} 
where this approach can be used it does not necessarily result in the most practical device, since {there can be particular} 
challenges and advantages associated with the new platform. 
For example, a spatial beamsplitter is passive and relatively easy to implement, whereas the equivalent transformation in frequency is more difficult to implement; {it} 
requires the use of active components {such as} 
electro-optic modulators (EOMs), or nonlinear processes such as Bragg four-wave mixing \cite{Lu_EOMs_PhysRevLett.120.030502, Agha:12}.

Developing a more systematic approach is challenging because the generation of multipartite states from conventional parametric sources requires measurement and postselection. If the ``full" state prior to any postselection were determined, the necessary configuration of the sources and linear components in the device could be determined using an input-output formalism, since the evolution up to the postselection is purely unitary. But it is not trivial in general to determine the {full} 
state that yields a target postselected state. 
Some progress in this direction has been made for devices in which one postselects on $N$-fold coincidences between detectors at the $N$ output modes identifying the individual qudits of an $N$-partite state \cite{Krenn_PhysRevLett.119.240403, Krenn_PhysRevX.11.031044, ruizgonzalez2022digital}.
This type of postselection is compatible with applications in which the qudits of the multipartite state need only be manipulated locally, {and is relevant for} 
communication and secret sharing protocols{. Indeed, }
recently a multipartite secret sharing protocol based on a postselected source of GHZ states was demonstrated \cite{QD_GHZme}. 

In this {paper} 
we 
present a {general} strategy for designing sources of multipartite states with this type of postselection. We will then apply this approach to the design of an efficient source of qutrit GHZ states, motivated by recent work on multipartite secret sharing, and anticipating a practical advantage with higher-dimensional systems for multipartite communication 
\cite{highdimQKD_https://doi.org/10.1002/qute.201900038}. {In Section \ref{section:notation} 
we introduce our notation, and we outline our approach for designing a source of the weakly squeezed state to which postselection will be applied.} We consider scenarios with and without restrictions on the types of components that can be implemented. {In Section \ref{section:graphs} 
we address the problem of determining the weakly squeezed state that corresponds to a target DV state after postselection;
we review a graph-based approach that can be used to this end.} 
{In Section \ref{section:example_GHZ} 
we apply this graph-based approach to a high-dimensional four-photon GHZ state, and in Section \ref{section:example_freqbin} we discuss the design of an integrated frequency-bin-encoded source of these states. We conclude in Section \ref{section:Conclusions}.} 



\section{Notation and Unitary Evolution}
\label{section:notation}

\begin{figure}
    \centering
    \includegraphics[width=0.48\textwidth]{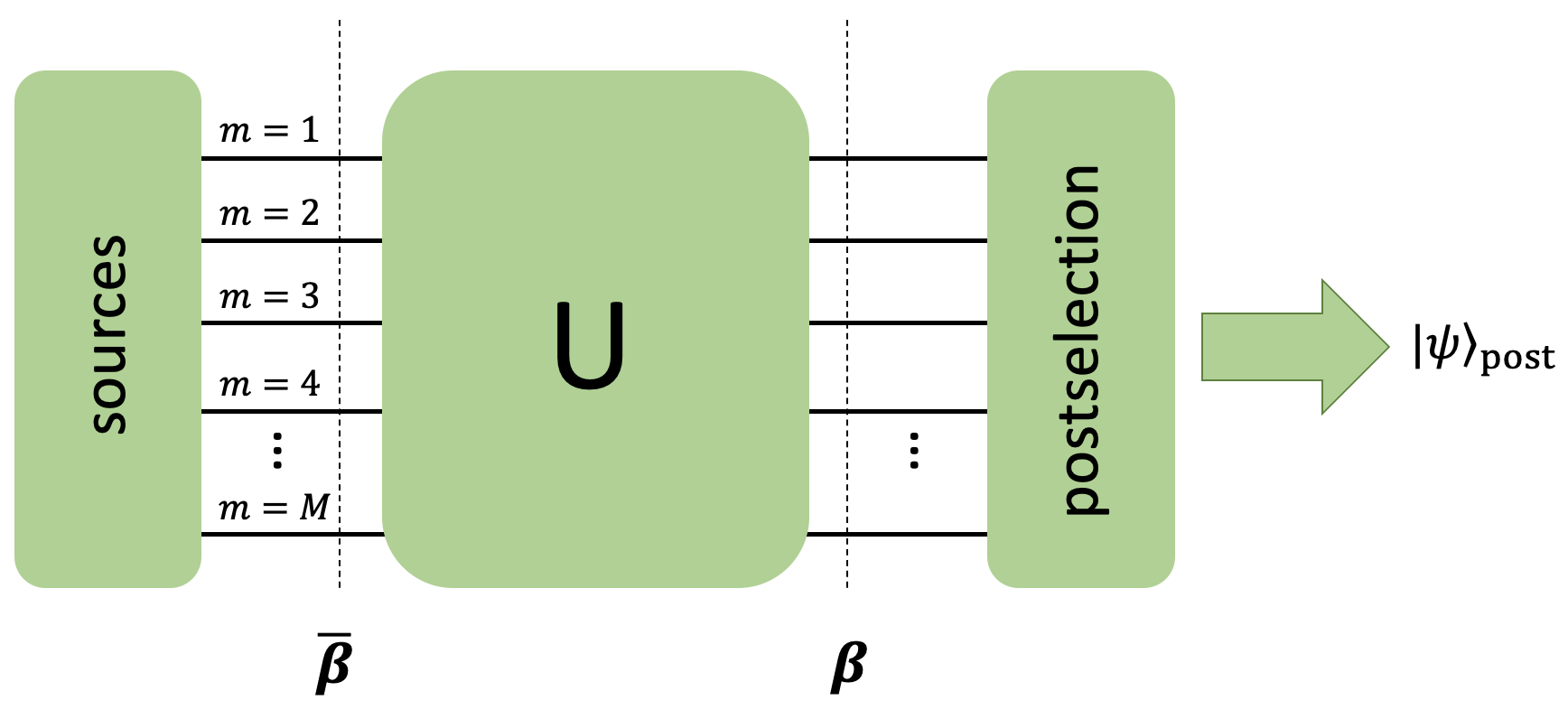}
    \caption{A sketch of the devices we consider. The box on the left represents a set of photon pair sources that collectively generate a multimode squeezed state; the different modes are represented by the black lines, and are labelled following the notation in Eq. \eqref{eq:C2_in_2}. The labelling of the modes is the same after the linear unitary circuit that is represented by the box labelled ``U". The box on the right represents measurement and postselection on the state, which results in a postselected state $\ket{\psi}_{\text{post}}$. We characterize the squeezed states before and after $\boldsymbol{U}$ in terms of the adjacency matrices $\boldsymbol{\bar{\beta}}$ and $\boldsymbol{\beta}$.}
    \label{fig:circuit}
\end{figure}

We consider devices involving photon pair sources, followed by linear evolution and postselection, as sketched in Fig. \ref{fig:circuit}. 
The wires in Fig. \ref{fig:circuit} represent the $M$ modes {that the generated photons can occupy.} 
We assume {that each photon pair source generates} 
{photons in a single Schmidt mode}. {Then} 
the sources in the device can be collectively described by a pair creation operator
\begin{align}
    C^{\dagger}_{II} = \sum_{m,m'}^{M} \overline{\beta}_{m,m'} a^{(in)\dagger}_{m} a^{(in)\dagger}_{m'}, \label{eq:C2_in_2}
\end{align}
provided that the sources are pumped coherently. The subscripts label the modes in which photons can be generated; these can be defined by more than one degree of freedom (DoF) {-- for example, path, polarization, frequency bin, etc.}

{When this type of device is used as a source of multipartite DV states, each mode $m$ encodes a logical state of a qudit, with {each group} of $d$ modes corresponding to a single qudit. It is useful to group the DoFs into those that label the qudits, and those that encode the logical states; we refer to these as ``external" and ``internal" DoFs, respectively. For example, in the first experiments involving GHZ states, path and polarization were the external and internal DoFs \cite{GHZ_PhysRevLett.82.1345}. Note that this distinction does not rely on the internal and external DoFs being physically different, but depends only on the encoding scheme. For example in an encoding scheme that involves only one physical DoF (such as path encoding), internal and external DoFs can still be identified by assigning to each mode two labels that index the qudits and the logical states.}


The $a^{(in)\dagger}_{m}$ are bosonic creation operators for the mode $m$, and $\overline{\beta}_{m,m'}$ is an amplitude associated with the creation of a photon pair in modes $m$ and $m'$. These amplitudes can be understood as the elements of a $M\times M$ symmetric matrix $\boldsymbol{\overline{\beta}}$ that describes the correlations between all the pairs of modes. 
{We define} $|\beta|^2 = \sum_{m,m'} |\overline{\beta}_{m,m'}|^2${; in the limit $|\beta|^2\ll 1$, this can be understood as the probability per pump pulse of generating a photon pair}.

The ``input" state to the subsequent unitary circuit can be written as 
\begin{align}
    \ket{\psi} = e^{C^{\dagger}_{II} - H.c.}\ket{\text{vac}} \label{eq:state_1},
\end{align}
neglecting time ordering effects. The generated photons then propagate through a linear unitary circuit. This can be described in a standard input-output picture as
\begin{align}
    a_{n}^{(out)} &= \sum_m U_{nm} a_m^{(in)}, \label{eq:in_out}
\end{align}
where we have defined a set of bosonic ``output" operators $a_{n}^{(out)}$, and the elements of the unitary are labelled $U_{nm}$. Since the unitary is linear, the state at the output of the unitary has the form of Eq. \eqref{eq:state_1}, now with
\begin{align}
    C^{\dagger}_{II} = \sum_{m,m'} {\beta}_{m,m'} a^{\dagger(out)}_m a^{\dagger(out)}_{m'}.  \label{eq:C2_out}
\end{align}
The output state is characterized by another matrix $\boldsymbol{\beta}$, related to $\boldsymbol{\overline{\beta}}$ by 
\begin{align}
    \boldsymbol{\beta} &= {\boldsymbol{U}} \overline{\boldsymbol{\beta}} {\boldsymbol{U}}^T, \label{eq:takagi?}
\end{align}
which can be found by writing Eq. \eqref{eq:C2_in} in terms of output mode operators using Eq. \eqref{eq:in_out} and comparing to Eq. \eqref{eq:C2_out}. 

If the target state at the output of the unitary is defined (that is, if $\boldsymbol{\beta}$ is known), Eq. \eqref{eq:takagi?} can be used to determine a set of sources and a unitary circuit that yield the correct {$\boldsymbol{\overline{\beta}}$} and $\boldsymbol{U}$. If one allows for an arbitrary unitary, {a solution can be identified} 
by taking the Takagi decomposition of $\boldsymbol{\beta}$. This yields a diagonal $\boldsymbol{\overline{\beta}}$, corresponding to a device with the simplest possible configuration for the photon-pair sources. For example, in a path encoding scheme, a diagonal $\boldsymbol{\overline{\beta}}$ would be implemented by having a degenerate photon pair source at the input of each path. The unitary resulting from the Takagi decomposition would be implemented by decomposing it into a mesh of beamsplitters \cite{Reck_PhysRevLett.73.58,Clements:16}, and in this way one could implement a source for the output state characterized by $\boldsymbol{\beta}$. 

Physically implementing the solution given by the Takagi decomposition generally requires that one be able to implement an arbitrary unitary transformation. In certain cases -- especially those in which an arbitrary unitary involves different types of transformations \cite{bouchard2024programmable} -- one may wish to put restrictions on the form of the unitary; for example, to avoid the use of difficult-to-implement or inefficient operations. One can seek solutions to Eq. \eqref{eq:takagi?} subject to restrictions on the form of $\boldsymbol{U}$, at the cost of requiring a more general form for $\overline{\boldsymbol{\beta}}$ than the diagonal matrix resulting from a Takagi decomposition. Accommodating this added complexity in $\overline{\boldsymbol{\beta}}$ requires sources that can generate non-degenerate photon pairs, which is relatively straightforward in many situations. 



{A less trivial task, and yet more interesting,} 
is solving Eq. \eqref{eq:takagi?} subject to the restriction that the unitary should involve no transformations of a subset of the relevant DoFs defining the modes 
$m$. For example, in Section \ref{section:example_freqbin} we will avoid frequency transformations in order to avoid the use of active components. We explicitly label the different DoFs that define the modes in Eq. \eqref{eq:C2_in_2}, writing 
\begin{align}
    C^{\dagger}_{II} = \sum_{n,n'}^{N} \sum_{\overline{n},\overline{n}'}^{\overline{N}} \overline{\beta}_{\overline{n}, n;\overline{n}', n'}a^{\dagger(in)}_{\overline{n},n} a^{\dagger(in)}_{\overline{n}',n'}. \label{eq:C2_in}
\end{align}
The barred and unbarred indices refer to DoFs that can and cannot be manipulated, respectively. By $N$ and $\overline{N}$ we denote the dimensionality of the two DoFs, and $N \overline{N} = M$. Even if the modes are labelled by more than two DoFs, these DoFs can still be grouped according to whether they can be manipulated by $\boldsymbol{U}$ or not. In this case the barred and/or unbarred indices in Eq. \eqref{eq:C2_in} would in turn be defined by more than one DoF, but the form of Eq. \eqref{eq:C2_in} and the following discussion still apply.

We partition $\boldsymbol{U}$ as 
\begin{align}
    \boldsymbol{U} = \begin{bmatrix}
        \boldsymbol{U}_{11} & \boldsymbol{U}_{12} & \cdots & \boldsymbol{U}_{1 {{N}}}\\
        \vdots & \vdots & \ddots & \vdots \\
        \boldsymbol{U}_{{{N}}1} & \boldsymbol{U}_{{{N}}2} & \cdots & \boldsymbol{U}_{{{N}} {{N}}} \label{eq:U_part}
    \end{bmatrix}
\end{align}
such that the $M\times M$ matrix is partitioned into ${{N}}^2$ blocks. Each block is a $\overline{N} \times \overline{N}$ matrix with the form
\begin{align}
    \boldsymbol{U}_{{n},{n}'} = \begin{bmatrix}
        U_{{1}{n};{1}{n}'} & U_{{1}{n};{2}{n}'} & \cdots & U_{{1}{n};\overline{N}{n}'} \\
        \vdots & \vdots & \ddots & \vdots \\
        U_{\overline{N}{n};{1}{n}'} & U_{\overline{N}{n};{2}{n}'} & \cdots & U_{\overline{N}{n};\overline{N}{n}'}
    \end{bmatrix}.
\end{align}
Since the unbarred indices refer to DoFs that cannot be manipulated by the unitary, {we require} 
$U_{\overline{n}n;\overline{n}'n'} = U_{\overline{n}n;\overline{n}'n'} \delta_{{n}{n}'}$; {that is,} 
$\boldsymbol{U}$ {must be} 
block-diagonal. 
Applying this restriction in Eq. \eqref{eq:takagi?} one finds the modified condition
\begin{align}
    \boldsymbol{\beta}_{ij} &= \boldsymbol{U}_{ii} \overline{\boldsymbol{\beta}}_{ij}\boldsymbol{U}^T_{jj}, \label{eq:block_takagi}
\end{align}
where $\boldsymbol{\beta}_{ij}$ and $\overline{\boldsymbol{\beta}}_{ij}$ are blocks of the matrices defined in Section \ref{section:notation}, partitioned in the same way as $\boldsymbol{U}$ in Eq. \eqref{eq:U_part}. The right hand side of Eq. \eqref{eq:block_takagi} is a matrix product, and no summation over the indices is implied.
One approach for finding a solution to Eq. \eqref{eq:block_takagi} is the following: 
\\
\begin{enumerate}
    \item {For the blocks} 
    $\boldsymbol{\beta}_{ij}$ {that are zero,} 
    the corresponding $\overline{\boldsymbol{\beta}}_{ij}$ {can be taken to be zero. The corresponding $\boldsymbol{U}_{ii}$ and $\boldsymbol{U}_{jj}$ remain unspecified.} 
    \item For diagonal nonzero blocks 
    ($i=j$), do the Takagi decomposition. This determines $\boldsymbol{\beta}_{ii}$ and the unitary $\boldsymbol{U}_{ii}$.
    \item For off-diagonal nonzero blocks ($i\neq j$) there are a few possibilities:
    \begin{enumerate}
        \item The relevant $\boldsymbol{U}_{ii}$ and $\boldsymbol{U}_{jj}$ are both determined by the diagonal blocks. Then $\overline{\boldsymbol{\beta}}_{ij}$ is obtained by matrix multiplication. 
        \item One of $\boldsymbol{U}_{ii}$ or $\boldsymbol{U}_{jj}$ is determined, but the other is undefined. In this case, one possibility might be to use the polar decomposition. For example, if $\boldsymbol{U}_{ii}$ is known, then we have $\boldsymbol{U}^{\dagger}_{ii} \boldsymbol{\beta}_{ij} = \overline{\boldsymbol{\beta}}_{ij} \boldsymbol{U}_{jj}^T$. The matrix on the left hand side can be calculated numerically and the polar decomposition can be done numerically.
        \item Neither $\boldsymbol{U}_{ii}$ nor $\boldsymbol{U}_{jj}$ are determined. In this case a singular value decomposition can be used. 
    \end{enumerate}
\end{enumerate}
In this way, one can find the $\boldsymbol{U}$ and $\boldsymbol{\overline{\beta}}$ that need to be implemented to produce the state represented by $\boldsymbol{\beta}$, subject to restrictions on the form of $\boldsymbol{U}$.




{With this approach, it is relatively straightforward to design the device once one determines the weakly squeezed state --characterized by $\boldsymbol{\beta}$ -- that yields the desired postselected state $\ket{\psi_{\text{post}}}$. 
It is simple (at least in principle) to determine the postselected state $\ket{\psi_{\text{post}}}$ resulting from a specified $\boldsymbol{\beta}$, but doing the reverse -- in particular, determining the $\boldsymbol{\beta}$ that results in a target $\ket{\psi_{\text{post}}}$ -- is more challenging. In general, the solution to the problem is not unique, and it is not even guaranteed that such a $\boldsymbol{\beta}$ exists for an arbitrary $\ket{\psi_{\text{post}}}$.} 



\begin{figure}[h]
    \centering
    \includegraphics[width=0.5\textwidth]{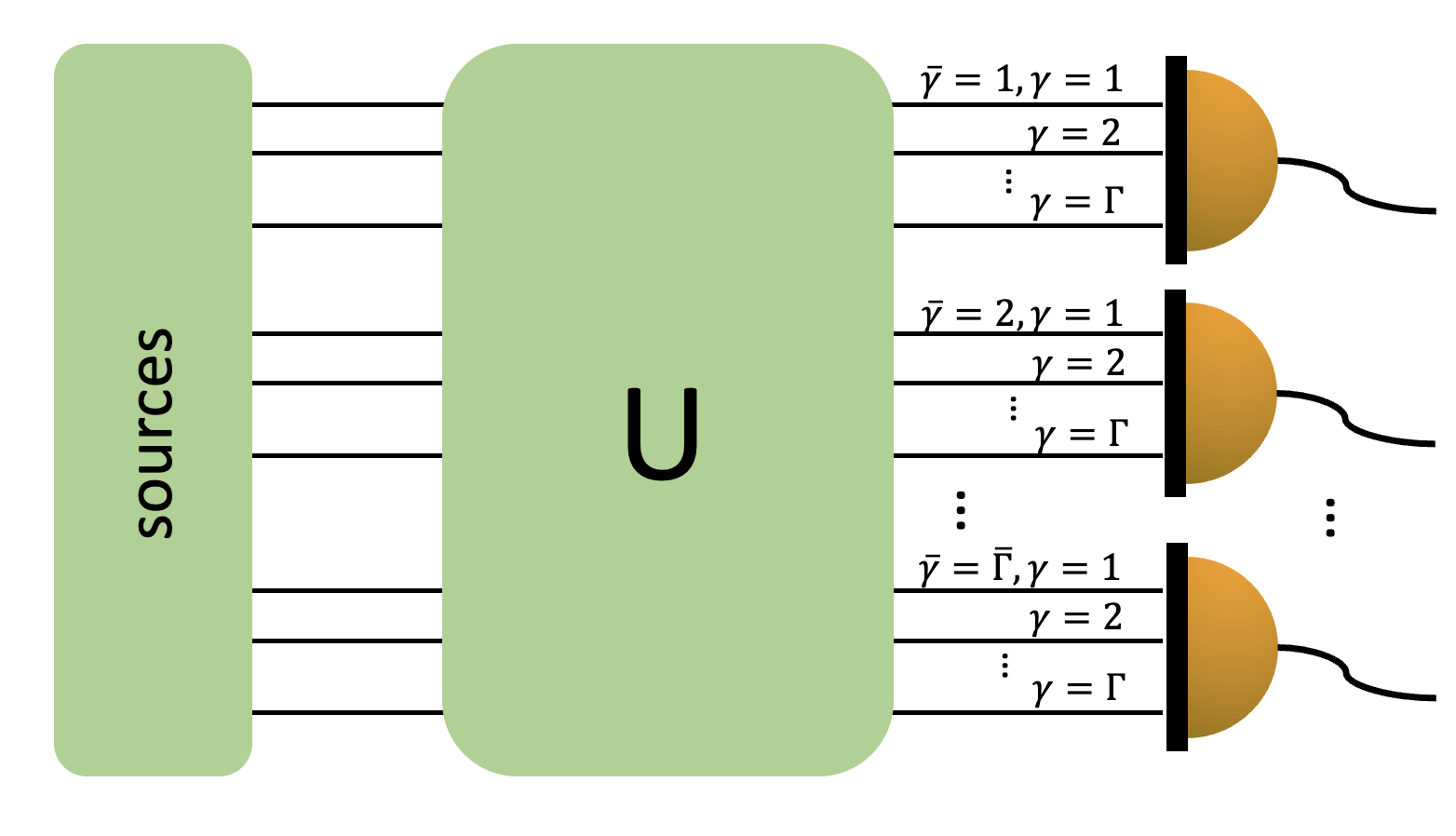}
    \caption{{The postselection scheme we assume in the remainder of this manuscript involves conditioning on detecting photons in each of the external modes, labelled by $\overline{\gamma}$; in the sketch this is represented by all the modes labelled by a particular value of $\overline{\gamma}$ leading to a single photon detector. The various modes leading to a single SPD are distinguished by the values of the internal DoF, which is labelled by the unbarred $\gamma$. The labelling of the modes is the same at the input of the unitary.}}
    \label{fig:postselection}
\end{figure}

{In the case in which the postselection involves detecting photons in each external mode (by which we mean the modes defined by the external DoFs; see Fig. \ref{fig:postselection}), this link can be made using the representation of weakly squeezed vacuum states as coloured weighted graphs. We focus on this type of postselection because this mapping between the postselected state and the weakly squeezed state is established, and also because it is compatible with near-term applications such as quantum secret sharing \cite{QD_GHZme}.  
One can envision more general scenarios with other postselection schemes, including those where only a subset of the external modes are measured, such that heralded DV states can be generated. These extensions would require establishing different ways to determine a $\boldsymbol{\beta}$ corresponding to $\ket{\psi_{\text{post}}}$, but the discussion in Section \ref{section:notation} applies very generally.}


\section{Graph representation}
\label{section:graphs}

We now review the graph representation of the state described by Eq. \eqref{eq:state_1} with Eq. \eqref{eq:C2_out}. We first disentangle the squeezing operator \cite{Ma_PhysRevA.41.4625} to rewrite Eq. \eqref{eq:state_1} as 
\begin{align}
    \ket{\psi} &= \mathcal{N} \text{exp}\left( \sum_{m,m'} w_{m,m'} a^{\dagger(out)}_{m} a^{\dagger(out)}_{m'} \right)\ket{\text{vac}} \label{eq:disent_1},
\end{align}
where $\mathcal{N}$ is a normalization constant \cite{Ma_PhysRevA.41.4625}.
Here again we identify the coefficients in Eq. \eqref{eq:disent_1} as the elements of a symmetric $M \times M$ matrix. In the low gain regime, where $\sum_{m,m'} |{\beta}_{m,m'}|^2 \ll 1$, the matrices $\boldsymbol{w}$ and $\boldsymbol{\beta}$ are approximately equivalent; more generally, the two are related as described {by Ma and Rhodes} 
\cite{Ma_PhysRevA.41.4625}.

The state in Eq. \eqref{eq:disent_1} can be represented as a coloured, weighted graph with $\Gamma$ vertices, and with the edges taking on $\overline{\Gamma}$ possible colours \cite{Krenn_PhysRevLett.119.240403}. The vertices represent the degrees of freedom used to label the different qudits (external DoFs -- recall the discussion below Eq. \eqref{eq:C2_in_2} ), while the colours represent the degrees of freedom that encode the qudits' logical state (internal DoFs). 
{It is useful to explicitly label each mode $m$ with two indices, which refer to the external and internal DoFs, as discussed below Eq. \eqref{eq:C2_in_2}. We write}
\begin{align}
    \ket{\psi} &= \mathcal{N} \text{exp}\left( \sum_{\overline{\gamma}, {\gamma},\overline{\gamma}', {\gamma}'} w_{\overline{\gamma}, \gamma; \overline{\gamma}', \gamma'} a^{\dagger(out)}_{\overline{\gamma},\gamma} a^{\dagger(out)}_{ \overline{\gamma}',\gamma'} \right)\ket{\text{vac}},\label{eq:disent_2}
\end{align}
where the barred and unbarred indices label the external and internal modes, respectively. {This notation is similar to Eq. \eqref{eq:C2_in}, and indeed the idea is the same -- namely, to divide the DoFs defining each mode into two groups, and to explicitly label these. However, the purpose of the grouping is different here than in Eq. \eqref{eq:C2_in}: Here it relates to the encoding scheme, where both logical states and different qudits need to be identified, whereas in Eq. \eqref{eq:C2_in}, the DoFs are grouped according to restrictions placed on the unitary implemented by the circuit, independent of the encoding scheme.} 

The weights of the edges of the graph representing $\ket{\psi}$ are given by the entries of the adjacency matrix $\boldsymbol{w}$; the element $w_{ \overline{\gamma},\gamma; \overline{\gamma}',\gamma'}$ gives the weight of the edge connecting the vertices ${\gamma}$ and ${\gamma}'$, with the colours $\overline{\gamma}$ and $\overline{\gamma}'$. We will arrange the entries in the adjacency matrix as 
\begin{align}
    \boldsymbol{w} = 
    \begin{bmatrix}
        \boldsymbol{w}_{00} & \boldsymbol{w}_{01} & \cdots & \boldsymbol{w}_{0{\Gamma}} \\
        \boldsymbol{w}_{10} & \boldsymbol{w}_{11} &\cdots & \boldsymbol{w}_{1{\Gamma}}\\
        \vdots & \vdots & \ddots & \vdots \\
        \boldsymbol{w}_{{\Gamma}0} & \boldsymbol{w}_{{\Gamma}1} & \cdots & \boldsymbol{w}_{{\Gamma}{\Gamma}}
    \end{bmatrix}, \label{eq:adjacency_blocks}
\end{align}
where each $\boldsymbol{w}_{{\gamma}{\gamma}'}$ in Eq. \eqref{eq:adjacency_blocks} is a $\overline{\Gamma} \times \overline{\Gamma}$ block of the matrix for a fixed pair of internal mode labels. That is, 
\begin{align}
    \boldsymbol{w}_{{\gamma}{\gamma}'} = \begin{bmatrix}
        \omega_{1{\gamma};1{\gamma}'} & \omega_{1{\gamma};2{\gamma}'} & \cdots & \omega_{1{\gamma};\overline{\Gamma}{\gamma}'} \\
        \omega_{2{\gamma};1{\gamma}'} & \omega_{2{\gamma};2{\gamma}'} & \cdots & \omega_{2{\gamma};\overline{\Gamma}{\gamma}'} \\
        \vdots & \vdots & \ddots & \vdots
        \\
        \omega_{\overline{\Gamma}\gamma;1{\gamma}'} & \omega_{\overline{\Gamma}{\gamma};2{\gamma}'} & \cdots & \omega_{\overline{\Gamma}{\gamma};\overline{\Gamma}{\gamma}'} .
    \end{bmatrix}
\end{align}
From Eq. \eqref{eq:disent_2} one sees that an edge on the graph can be understood as representing an amplitude associated with generated photon pairs being distributed in the modes $\{\overline{\gamma}, {\gamma}\}$ and $\{\overline{\gamma}',\gamma'\}$ \cite{Krenn_PhysRevLett.119.240403,ruizgonzalez2022digital}.

Eq. \eqref{eq:disent_2} can be written as the expansion
\begin{align}
    \ket{\psi} 
    &= \nonumber \mathcal{N} \bigg( 1 + \sum_{\overline{\gamma},\gamma,\overline{\gamma}',\gamma'} w_{\overline{\gamma},\gamma;\overline{\gamma}',\gamma'} a^{\dagger(out)}_{\overline{\gamma},\gamma} a^{\dagger(out)}_{\overline{\gamma}',\gamma'}\\ &+ \nonumber \frac{1}{2} \sum_{\substack{\overline{\gamma},\gamma,\overline{\gamma},\gamma'\\
    \overline{\delta},\delta,\overline{\delta}',\delta'}} w_{\overline{\gamma},\gamma;\overline{\gamma}',\gamma'} w_{\overline{\delta},\delta;\overline{\delta}',\delta'} \\ &\times a^{\dagger(out)}_{\overline{\gamma},\gamma,} a^{\dagger(out)}_{\overline{\gamma}',\gamma'} a^{\dagger(out)}_{\overline{\delta},\delta} a^{\dagger(out)}_{\overline{\delta}',\delta',} + ...\bigg) \ket{\text{vac}} \label{eq:state_big_main},
\end{align}
which is a superposition of vacuum, two-photon terms, four-photon terms, and so on. We assume a regime where $w_{\overline{\gamma},\gamma;\overline{\gamma}',\gamma'}\ll1$, such that one can neglect terms beyond a certain order to good approximation. The order to which we expand depends on the number of ``output modes" $\overline{\Gamma}$ we consider; for example, we will see that for the GHZ example (Fig. \ref{fig:GHZ_graph}) -- where $\overline{\Gamma}=4$ and the graph has four vertices -- it is sufficient to work up to the four-photon terms (order $w^2$).

One can think of the postselection as eliminating certain terms in the full state, while leaving those that correspond to some target state. By postselecting on coincidences across all the output modes, we effectively generate a state proportional to the terms of Eq. \eqref{eq:state_big_main} in which all the output mode labels appear at least once (see Appendix \ref{appendix:perfectmatch}). The lowest-order terms appearing in the postselected state can be understood in terms of the perfect matchings of the graph \cite{Krenn_PhysRevLett.119.240403} {-- that is, the sets of edges such that each vertex in the graph is connected to exactly one edge.}

Therefore if one has target DV state to generate with postselection, a source can be designed by finding a graph with perfect matchings corresponding to the terms in this target state. This can be done using methods that have been developed by others \cite{Krenn_PhysRevX.11.031044,ruizgonzalez2022digital}. Once this graph is known, the weights of the edges can be used to find the form of $\boldsymbol{\beta}$, which characterizes the state before postselection (recall Eq. \eqref{eq:C2_out}). With this and Eq. \eqref{eq:takagi?} one can find the necessary configuration for the unitary and photon pair sources; with this, a source for the target state is fully specified.

\section{Example: High-dimensional GHZ state}
\label{section:example_GHZ}

We take our target state to be a high-dimensional four-photon GHZ state, with the from 
\begin{align}
    \ket{\psi} = \frac{1}{\sqrt{3}} \left( \ket{0000} + \ket{1111} + \ket{2222} \right). \label{eq:GHZ_state}
\end{align}
GHZ states are resources in many applications, including some that are compatible with postselection on all the photons \cite{PhysRevLett.115.020502,PhysRevA.59.1829_GHZ_secret_sharing,PhysRevA.63.054301_GHZ_dense_coding}. For example, a multipartite secret sharing protocol was recently demonstrated using single-photon sources and postselection \cite{QD_GHZme}. The generation of postselected qubit GHZ states has been discussed extensively \cite{PhysRevLett.86.4435,Silberhorn_PhysRevLett.129.150501,Lo2023,PhysRevA.107.033714,PhysRevLett.132.130604}, but the generation of photonic qudit GHZ states is less well-understood; we address this example as an extension on this earlier work. These types of sources are also motivated by expected improvements to the efficiency of communication protocols when when implemented with high-dimensional states \cite{highdimQKD_https://doi.org/10.1002/qute.201900038}. 

\begin{figure}
    \centering
    \includegraphics[width=0.5\textwidth]{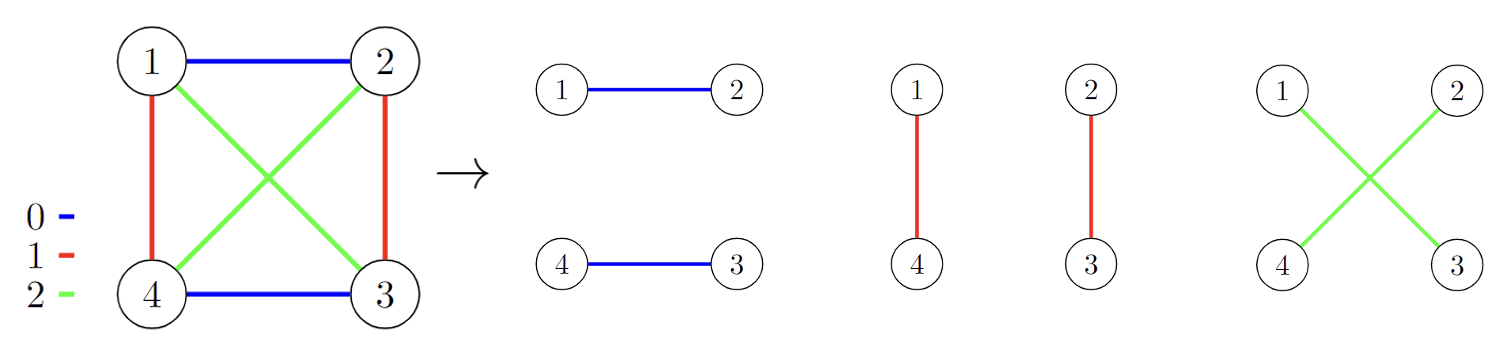}
    \caption{A graph representing a weakly-squeezed multimode state (left), and its three perfect matchings (right). The perfect matchings correspond to a qutrit GHZ state {\cite{ruizgonzalez2022digital}}.}
    \label{fig:GHZ_graph}
\end{figure}

The graph with the relevant perfect matchings is given in Fig. \ref{fig:GHZ_graph}. We point out that the graph for a qubit GHZ state can be obtained from this by setting the weights of the green edges to zero, leaving only two perfect matchings corresponding to the two terms in the qubit GHZ state. We also point out that a four-photon GHZ state with higher dimensionality than the one we consider here requires the use of ancilla photons; the additional perfect matchings that would be required to increased the dimensionality of the state cannot be accommodated with a four-vertex graph, without introducing unwanted crossterms \cite{ruizgonzalez2022digital}
. We restrict ourselves to the four-photon case here; the approach for designing sources of states with a higher number of qudits is the same, but the implementation of these larger devices is more challenging. 

The graph's adjacency matrix can be written as
\begin{align}
    \boldsymbol{w} = 
    \begin{bmatrix}
        \boldsymbol{w}_{00} & \boldsymbol{w}_{01} & \boldsymbol{w}_{02} \\
        \boldsymbol{w}_{10} & \boldsymbol{w}_{11} & \boldsymbol{w}_{12} \\
        \boldsymbol{w}_{20} & \boldsymbol{w}_{21} & \boldsymbol{w}_{22} \\
    \end{bmatrix}, \label{eq:GHZ_adjacency_blocks}
\end{align}
with 
\begin{align}
    \boldsymbol{w}_{00} &= 
    \begin{bmatrix}
        0 & w_{1,0;2,0} & 0 & 0 \\
        w_{1,0;2,0} & 0 & 0 & 0 \\
        0 & 0 & 0 & w_{3,0;4,0} \\
        0 & 0 & w_{3,0;4,0} & 0 
    \end{bmatrix}, \\
    \boldsymbol{w}_{11} &= 
    \begin{bmatrix}
        0 & 0 & 0 & w_{1,1;4,1} \\
        0 & 0 & w_{2,1;3,1} & 0 \\
        0 & w_{2,1;3,1} & 0 & 0 \\
        w_{1,1;4,1} & 0 & 0 & 0 
    \end{bmatrix},\\
    \boldsymbol{w}_{22} &= 
    \begin{bmatrix}
        0 & 0 & w_{1,2;3,2} & 0 \\
        0 & 0 & 0 & w_{2,2;4,2}  \\
        w_{1,2;3,2} & 0 & 0 & 0 \\
        0 & w_{2,2;4,2} & 0 & 0 
    \end{bmatrix}, \label{eq:adj_block}
\end{align}
where we have used 
$w_{\overline{\gamma},\gamma; \overline{\gamma}',\gamma' } = w_{\overline{\gamma}', \gamma';  \overline{\gamma},\gamma}$. All the off-diagonal blocks in Eq. \eqref{eq:GHZ_adjacency_blocks} are zero, since the graph features no two-coloured edges. The adjacency matrix is specified by six unique weights, representing the graph's six edges.

So far we have left the weights of the edges unspecified; the relative magnitudes and phases of the terms in Eq. \eqref{eq:GHZ_state} can be modified by adjusting the elements of $\boldsymbol{w}$. For the ``balanced" state written in Eq. \eqref{eq:GHZ_state}, the weights should all be set equal. {One can define the weights relative to any constant (for example, unity tends to be convenient) when doing the decomposition required to solve Eq. \eqref{eq:takagi?}}; in practice, the weights should be much smaller than unity to satisfy the assumption of a low-gain regime. The $\boldsymbol{\overline{\beta}}$ resulting from the matrix decomposition can be rescaled by a constant factor to ensure this. 

Once the weights are specified, Eq. \eqref{eq:takagi?} can be solved numerically. The simplest approach is to interpret Eq. \eqref{eq:takagi?} as the Takagi decomposition of the matrix $\boldsymbol{\beta}$. A physical implementation for the resulting $\boldsymbol{U}$ can be found by decomposing it into a product of two-mode unitaries \cite{Reck_PhysRevLett.73.58,Clements:16}. This mesh of two-mode unitaries can then be implemented using whatever components are relevant for the encoding scheme and physical devices with which one chooses to work \cite{universallinear_doi:10.1126/science.aab3642, Lukens:17, bouchard2024programmable}. 

\section{Implementation in frequency bin encoding on a passive chip}
\label{section:example_freqbin}

We now outline some details about the implementation of photon pair sources and unitary transformations in a frequency bin encoding scheme in integrated devices. 

\subsection{Sources of frequency-bin-entangled photon pairs}
\label{section:sources}

\begin{figure}[h]
    \centering
    \subfloat[]{
        \includegraphics[width=0.45\textwidth]{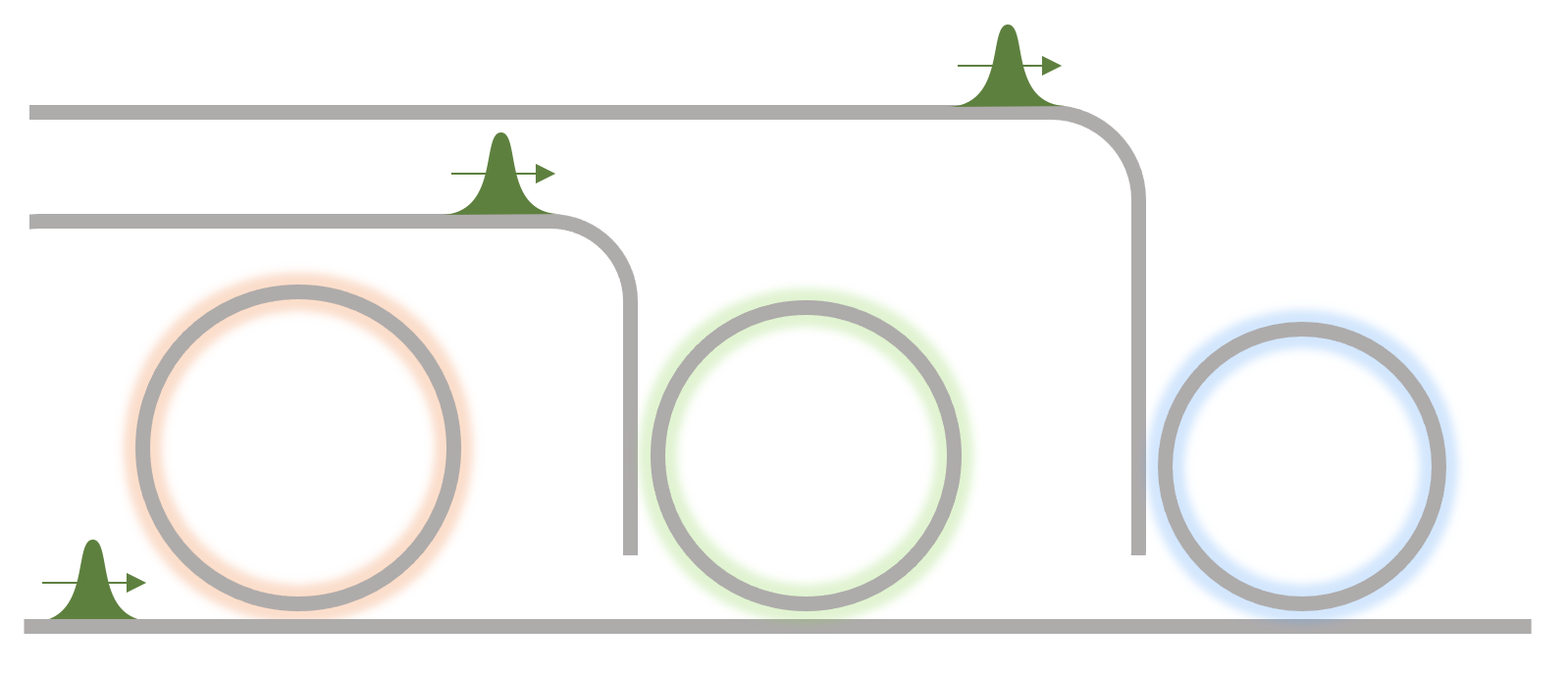}
    }
    
    \subfloat[]{
        \includegraphics[width=0.45\textwidth]{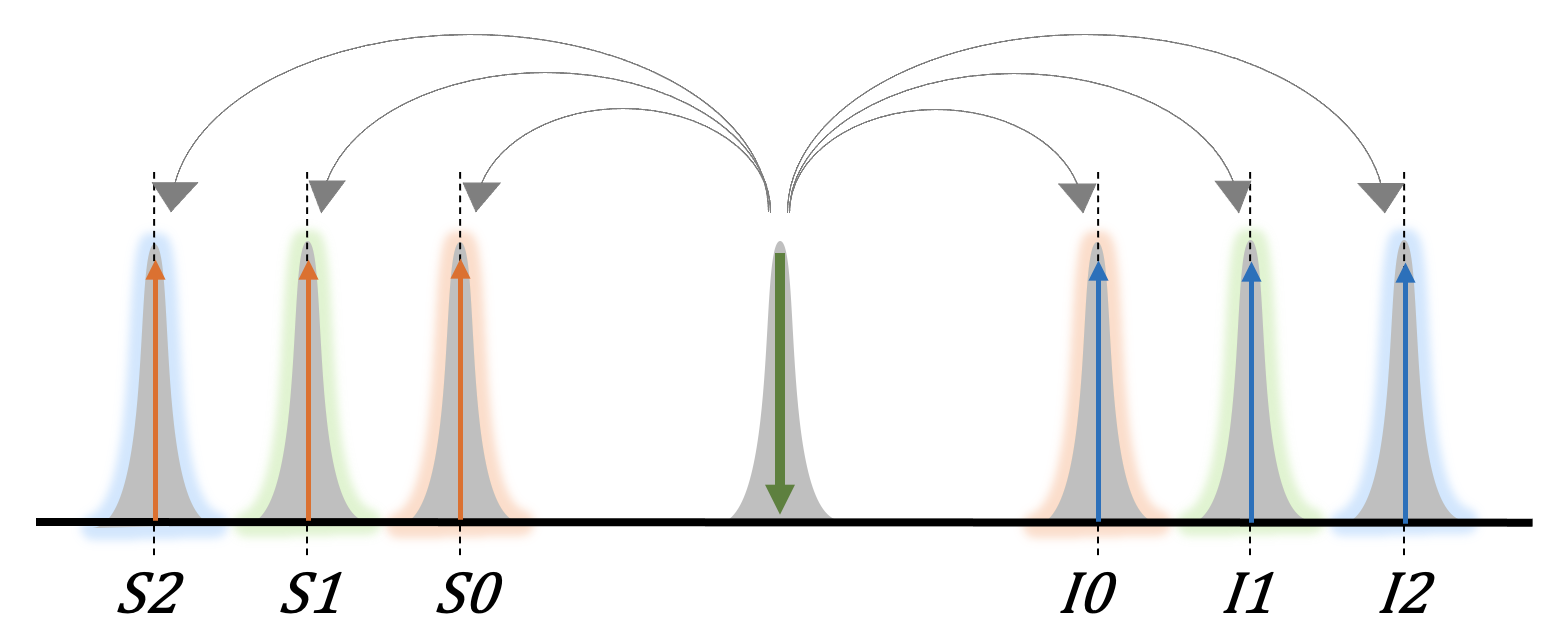}
    }
    \caption{A source of frequency-bin-encoded photon pairs (a), and a possible configuration for the pump and generated photons (b). Pairs of photons can be generated in different pairs of ring resonances, labelled {$S0$, $S1$, $S2$, $I0$, $I1$, $I2$}. Different resonances in (b) belong to physically different rings in (a), as indicated by the highlights.}
    \label{fig:single_pump}
\end{figure}

Frequency-bin-entangled photon pairs can be generated through spontaneous four-wave mixing (SFWM) in microring resonators. This can be done by generating photon pairs across a comb of resonances, as sketched in Fig. \ref{fig:single_pump}. These types of sources can be implemented using a single low-FSR resonator, or using several microrings driven coherently, such that the resonances in the comb belong to different physical components. We focus on the latter approach, which can be more efficient than the former by avoiding tradeoffs that arise for a single ring \cite{Liscidini:19}. Moreover, this approach is more flexible; one independently tune the resonances and modulate the pump fields for each ring. This tuneability is particularly important if one aims to design passive sources.

{We adopt an encoding scheme that involves the photons' paths and frequencies. Each resonance frequency (see Fig. \ref{fig:single_pump}) has two labels associated with it: One refers to its ``frequency range" ($S$ or $I$), and the other refers to the frequency bin (labelled $0,1,$, and so on). The frequency bin is defined with respect to some reference frequency -- in this case, the central pump frequency -- regardless of whether it is red- or blue-detuned from this reference frequency {(see Fig. \ref{fig:single_pump})}. 
Here the path (labelled by lowercase letters $\mathsf{a},\mathsf{b},...$) and ``frequency range" (labelled $S$ or $I$) are the external DoFs, 
which are represented by barred indices in Eq. \eqref{eq:disent_2}; the ``frequency bin" is the external DoF states (unbarred indices in Eq. \eqref{eq:disent_2}).}

One approach to designing a source of postselected frequency-bin-encoded states would be to use the Takagi decomposition of the matrix $\boldsymbol{\beta}$ for the target state (as discussed in Section \ref{section:notation}). The resulting diagonal $\overline{\boldsymbol{\beta}}$ would be implemented by having a photon pair source at the input of each path, generating photon pairs in the relevant frequency bins as sketched in Fig. \ref{fig:single_pump}. The corresponding $\boldsymbol{U}$ could be implemented using a mesh of EOMs and passive linear components, such as directional couplers (DCs) and add-drop microrings 
\cite{Lukens:17,Reck_PhysRevLett.73.58}. In this way one could construct a device that can implement an arbitrary $\boldsymbol{\beta}$, similar to work that has been done in path encoding. 

But EOMs can be lossy and challenging to integrate, especially compared to passive components. Although they would be necessary to implement an arbitrary unitary, analogous to implementations in path encoding \cite{Bao2023, Clements:16, Reck_PhysRevLett.73.58}, they are not necessarily needed in sources of particular classes of states. In some cases, the requirement for EOMs can be relaxed by configuring the sources appropriately, as discussed in Section \ref{section:notation}. For example, this can entail having sources that generate nondegenerate photon pairs, which is a straightforward generalization for the types of sources on which we focus \cite{Clementi2023}. 

\subsection{GHZ state}

\begin{figure}[h]
    \centering
    \subfloat[]{
        \includegraphics[width=0.45\textwidth]{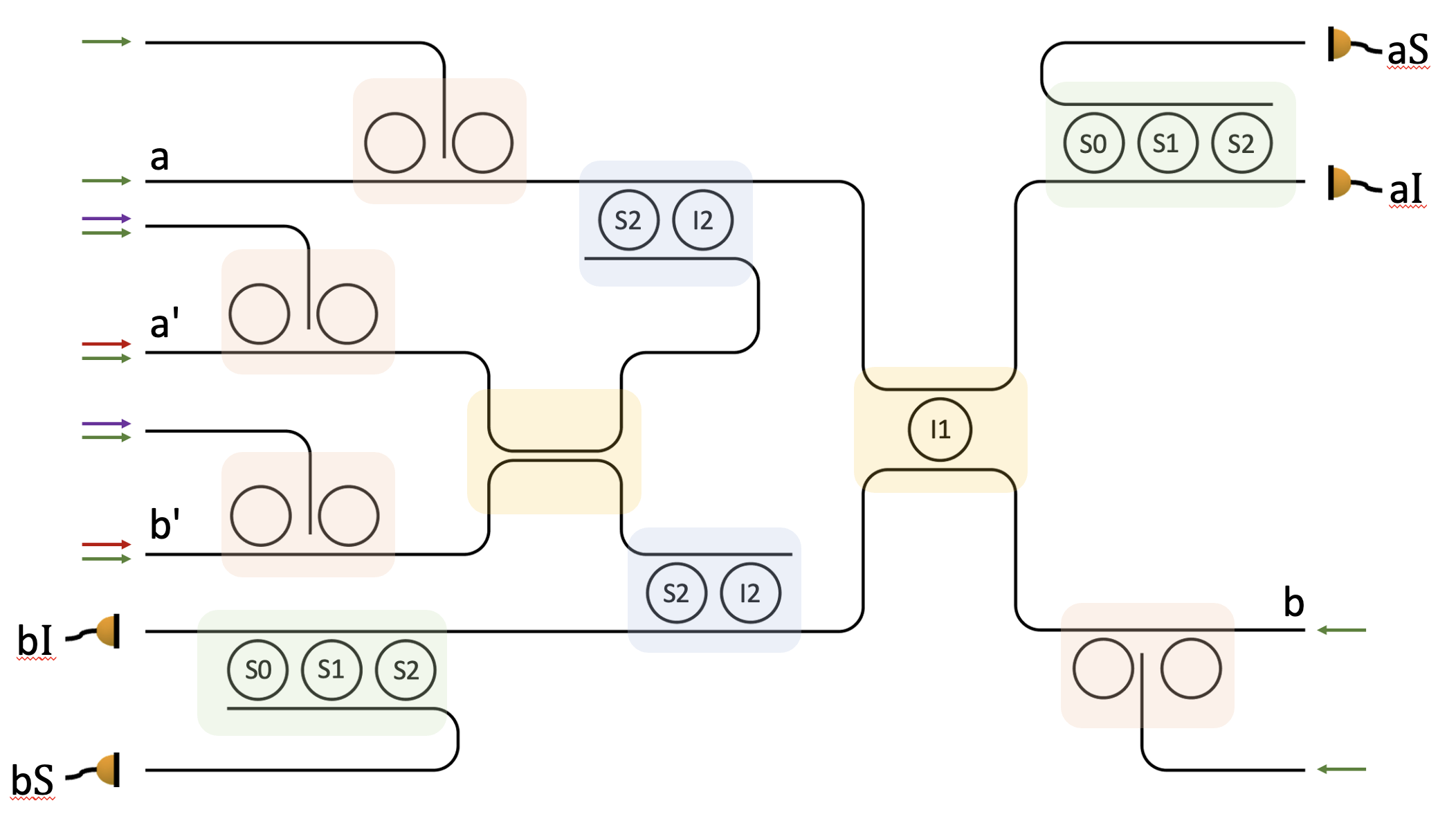}
    }

    \subfloat[]{
        \includegraphics[width=0.45\textwidth]{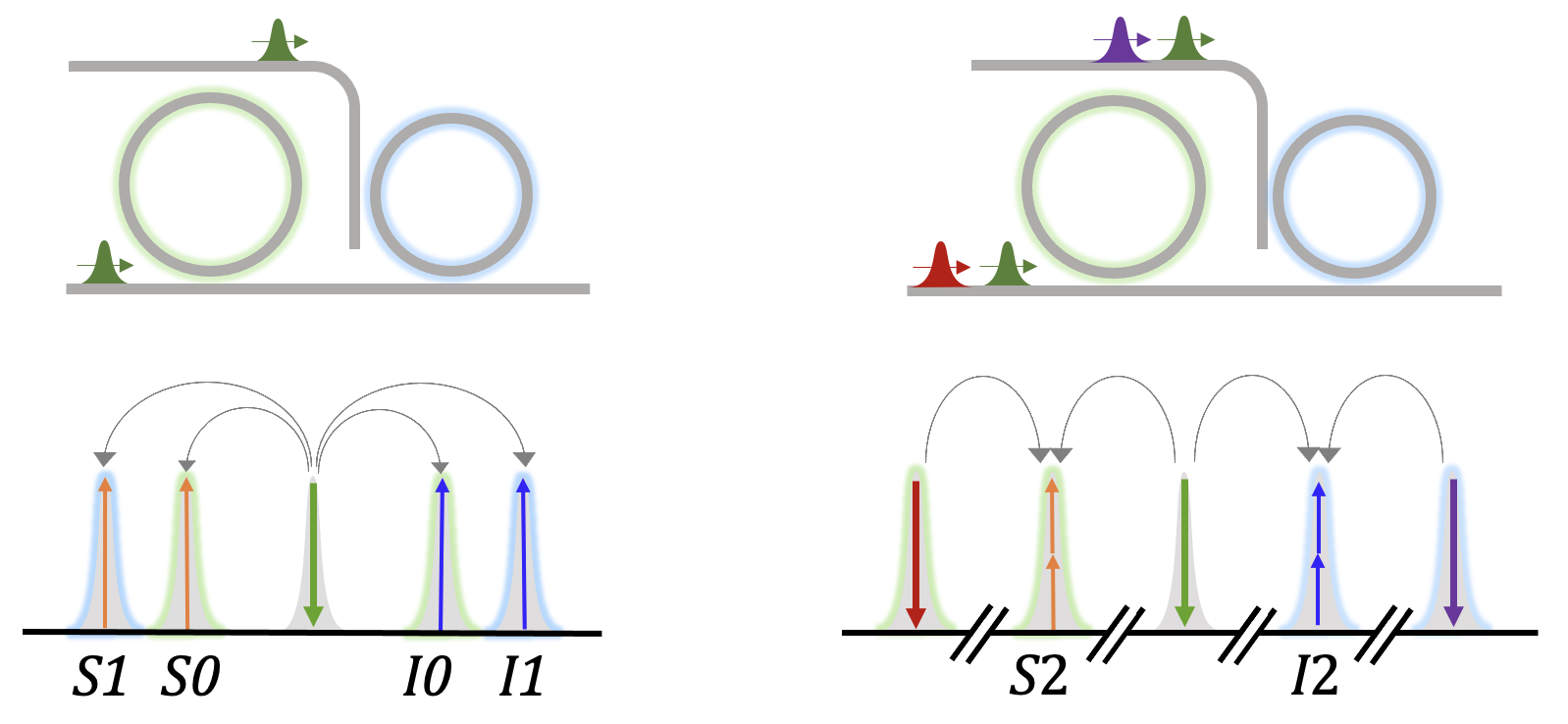}
    }
    \caption{Sketch of a passive integrated source of postselected qutrit GHZ states (a). The arrows indicate the input pump fields, and the detectors indicate output ports. The sources are highlighted in orange; the configurations of these are indicated in panel (b). The linear components are highlighted in yellow, blue, and green.}
    \label{fig:ghz_device}
\end{figure}

As mentioned above, we adopt an encoding scheme in which the four output modes -- indexing the four qutrits of the GHZ state -- are labelled by a path ($\mathsf{a}$ or $\mathsf{b}$) and a frequency range (S or I). We take the vertices numbered in Fig. \ref{fig:GHZ_graph} to be labelled by these two DoFs as follows: 
\begin{align}
    1 &\rightarrow \mathsf{a}S, \\
    2 &\rightarrow \mathsf{a}I,\\ 
    3 &\rightarrow \mathsf{b}S,\\
    4 &\rightarrow \mathsf{b}I.
\end{align}
The logical state is determined by the frequency bin (0, 1, or 2). If we restrict the form of $\boldsymbol{U}$ to involve no manipulation of frequency, then we partition the matrices into 36 blocks. For example, we write 
        %
        %
        %
        %
        %
        %
%
\begin{align}
    \boldsymbol{\beta} = \begin{bmatrix}
        \boldsymbol{0} & \boldsymbol{0} & \boldsymbol{0} & \boldsymbol{\beta}_{S0,I0} & \boldsymbol{0} & \boldsymbol{0} \\
        \boldsymbol{0} & \boldsymbol{0} & \boldsymbol{0} & \boldsymbol{0} & \boldsymbol{\beta}_{S1,I1} & \boldsymbol{0} \\
        \boldsymbol{0} & \boldsymbol{0} & \boldsymbol{\beta}_{S2,S2} & \boldsymbol{0} & \boldsymbol{0} & \boldsymbol{0} \\
        \boldsymbol{\beta}_{I0,S0} & \boldsymbol{0} & \boldsymbol{0} & \boldsymbol{0} & \boldsymbol{0} & \boldsymbol{0} \\
        \boldsymbol{0} & \boldsymbol{\beta}_{I1,S1} & \boldsymbol{0} & \boldsymbol{0} & \boldsymbol{0} & \boldsymbol{0} \\
       \boldsymbol{0} & \boldsymbol{0} & \boldsymbol{0} & \boldsymbol{0} & \boldsymbol{0} & \boldsymbol{\beta}_{I2,I2}
    \end{bmatrix}, 
\end{align}
where we have identified the blocks that are zero based on the graph in Fig. \ref{fig:GHZ_graph}. Each $\boldsymbol{\beta}_{Ji,J'i'}$ is the block of $\boldsymbol{\beta}$ is a $2\times2$ containing the amplitudes associated with photon pairs being generated with one photon in the $i^{\text{th}}$ frequency bin within the frequency range $J$, and the other in bin $i'$ within the frequency range $J'$. With reference to Fig. \ref{fig:GHZ_graph} --{and the corresponding adjancency matrix given in Eqs. \eqref{eq:GHZ_adjacency_blocks} to \eqref{eq:adj_block}} -- we can identify
\begin{align}
    \label{eq:beta_1} \boldsymbol{\beta}_{S0,I0} &= \begin{bmatrix}
        \beta_{\mathsf{a}S0,\mathsf{a}I0} & 0 \\
        0 & \beta_{\mathsf{b}S0,\mathsf{b}I0} 
    \end{bmatrix} = \boldsymbol{\beta}_{I0,S0}^T, \\
    \label{eq:beta_2} \boldsymbol{\beta}_{S1,I1} &= \begin{bmatrix}
        0 & \beta_{\mathsf{a}S1,\mathsf{b}I1} \\
        \beta_{\mathsf{b}S1,\mathsf{a}I1} & 0 
    \end{bmatrix} = \boldsymbol{\beta}_{I1,S1}^T, \\
    \label{eq:beta_3} \boldsymbol{\beta}_{S2,S2} &= \begin{bmatrix}
        0 & \beta_{\mathsf{a}S2,\mathsf{b}S2} \\
        \beta_{\mathsf{b}S2,\mathsf{a}S2} & 0 
    \end{bmatrix}, \\
    \label{eq:beta_4} \boldsymbol{\beta}_{I2,I2} &= \begin{bmatrix}
        0 & \beta_{\mathsf{a}I2,\mathsf{b}I2} \\
        \beta_{\mathsf{b}I2,\mathsf{a}I2} & 0 
    \end{bmatrix}.
\end{align}
For a balanced GHZ state, all the amplitudes above should have equal magnitude. We will take the relative phases to be zero, so we set all the nonzero amplitudes in Eqs. \eqref{eq:beta_1} -- \eqref{eq:beta_4} to unity. 
\begin{align}
    \boldsymbol{\beta}_{S0,I0} &= \begin{bmatrix}
        1 & 0 \\
        0 & 1 
    \end{bmatrix} = \boldsymbol{\beta}_{I0,S0}^T, \label{eq:beta_S0I0} \\
    \boldsymbol{\beta}_{S1,I1} &= \begin{bmatrix}
        0 & 1 \\
        1 & 0 
    \end{bmatrix} = \boldsymbol{\beta}_{I1,S1}^T = \boldsymbol{\beta}_{S2,S2} = \boldsymbol{\beta}_{I2,I2}.
\end{align}
The Takagi decomposition for the diagonal blocks gives 
\begin{align}
    \boldsymbol{U}_{S2,S2} = \boldsymbol{U}_{I2,I2} &= \frac{1}{\sqrt{2}}  \begin{bmatrix}
        i & 1 \\ -i & 1
    \end{bmatrix}. \label{eq:U_S2S2} 
\end{align}
The singular value decomposition of $\boldsymbol{\beta}_{S1,I1}$ yields 
\begin{align}
    \boldsymbol{U}_{S1,S1} &=\mathbb{1} \\
    \boldsymbol{U}_{I1,I1} &=\begin{bmatrix}
        0 & 1 \\ 1 & 0
    \end{bmatrix} \label{eq:U_I1I1}
\end{align}
and since Eq. \eqref{eq:beta_S0I0} is already diagonal we have $\boldsymbol{U}_{S0,S0} = \boldsymbol{U}_{I0,I0} = \mathbb{1}$. From these decompositions we also have 
\begin{align}
    \overline{\boldsymbol{\beta}}_{S0,I0} = \overline{\boldsymbol{\beta}}_{I0,S0} = \overline{\boldsymbol{\beta}}_{S1,I1} = \overline{\boldsymbol{\beta}}_{I1,S1} = \overline{\boldsymbol{\beta}}_{S2,S2} = \overline{\boldsymbol{\beta}}_{I2,I2}  &=\mathbb{1}. \label{eq:GHZ_betabar}
\end{align}
%
The photonic device implementing this $\boldsymbol{U}$ and $\boldsymbol{\overline{\beta}}$ is sketched in Fig. \ref{fig:ghz_device}.

The microrings highlighted in orange are the sources tuned to generate the weak multimode squeezed state characterized by $\boldsymbol{\overline{\beta}}$, as specified by Eq. \eqref{eq:GHZ_betabar}. Fig. \ref{fig:ghz_device}b illustrates the pump and resonance configurations for the sources. Here the blocks of $\boldsymbol{\overline{\beta}}$ are all diagonal, so the ring resonances and pump fields are tuned to generate photon pairs in the same frequency bin, but more general configurations can easily be implemented when necessary \cite{Clementi2023} -- an example of a source that requires this is given in Appendix \ref{appendix:L}. 

The DC highlighted in yellow implements the unitaries in Eq. \eqref{eq:U_S2S2}, when properly tuned \cite{Reck_PhysRevLett.73.58}. The microrings labelled $S2$ and $I2$ are add-drop rings with resonance frequencies at $S2$ and $I2$ respectively, and no other relevant frequency; their purpose is to route the photons into the output ports (such that the path we label $\mathsf{a}'$ and $\mathsf{b}'$ are effectively the same as paths $\mathsf{a}$ and $\mathsf{b}$. Similarly, the microring labelled $I1$ is an add-drop ring resonant only with the frequency $I1$, but its role is to implements Eq. \eqref{eq:U_I1I1} with no effect on the other frequencies. Finally, the rings labelled $S0$, $S1$, $S2$ before each set of output ports are add-drop rings acting as demultiplexers; the demultiplexing could be done off-chip instead.

The GHZ generation rates for this source are the same as we expect for a similar source of qubit GHZ states. {As in our earlier work \cite{banic_PhysRevA.109.013505}, we envision the source in Fig. \ref{fig:ghz_device} to be a pair of silicon microring resonators. The generation of nearly uncorrelated photon pairs with $|\beta|^2\approx 0.1$ has been demonstrated in these sources, driven by picojoule pump pulses with $\sim$10 ps durations, and a 10 MHz repetition rate
 \cite{Grassani2016}. For these sources and pumping scheme, we expect an on-chip GHZ generation rate of $10^4 - 10^5$ Hz \cite{banic_PhysRevA.109.013505}.

\section{Conclusion}
\label{section:Conclusions}

We have described a strategy for designing sources of postselected multipartite states. The strategy has two steps: First, the state just prior to the postselection is determined given a target state and postselection scheme. Second, the necessary configuration sources and linear components needed in the device is determined by solving Eq. \eqref{eq:takagi?}, with the option of imposing constraints on the form of the circuit. The approach we lay out for the second step is completely general, but our approach for the first step is not; we use the results of a graph-based approach that applies to the postselection scheme in which all the qudits are measured. 

We use this approach to design an integrated source of frequency-bin-encoded high-dimensional GHZ states. We do all this without reference to any existing schemes, and taking into account the challenges in this specific platform. In particular, we set as a constraint that the device should avoid the use of electro-optic modulators; the resulting passive device result in an on-chip GHZ generation rate on the order of 10 kHz for picojoule pump pulses. 

Our approach can be applied directly to other resources -- regardless of the encoding scheme and physical implementation -- for applications that are compatible with this postselection scheme. {Furthermore, graph-based solutions for \emph{heralded} Bell states and GHZ states have been identified \cite{Krenn_PhysRevX.11.031044}; sources of these heralded states could easily be designed using our approach, although the practical implementation of these sources would be more difficult due to the relatively many ancilla photons and detectors required. An obvious direction for future work is to explore other types of postselection more rigorously, and to seek strategies to reduce the experimental resources required to generate heralded states.} 

\begin{acknowledgments}
 M.B. acknowledges support from the Quantum Research and Development Initiative, led by the National Research Council Canada, under the National Quantum Strategy. M.L. acknowledges support by PNRR MUR project PE0000023-NQSTI. J.E.S. acknowledges support from the Natural Sciences and Engineering Research Council of Canada. 
\end{acknowledgments}

\bibliography{apssamp}

\onecolumngrid
\appendix
\pagebreak

\section{Understanding the postselected state in terms of perfect matchings}
\label{appendix:perfectmatch}

Here we elaborate on the link between the state produced by applying measurement and postselection on the weakly squeezed state represented by the adjacency matrix $\boldsymbol{w}$, and the perfect matchings of the graph represented by that adjacency matrix. 

In Eq. \eqref{eq:state_big_main} (see Section \ref{section:graphs}) we write the multimode squeezed state as an expansion in powers of the weights $w_{m,m'}$. The terms in the expansion can be understood with reference to the graph that corresponds to $\boldsymbol{\beta}$. For example, each two-photon term in Eq. \eqref{eq:state_big_main} corresponds to an individual edge of the graph, where the term describes the generation of a photon in the modes represented by the colours of the edge, and the vertices it connects (c.f. Section \ref{section:graphs}). Similarly, each four-photon term can be understood in terms of a pair of edges in the full graph. 


As an example, consider the state represented by the graph in Fig. \ref{fig:GHZ_graph_app}
\begin{figure}[h]
    \centering
    \begin{tikzpicture}[scale=0.7]
  \node[circle,draw] (1) at (0,0) {1};
  \node[circle,draw] (2) at (3,0) {2};
  \node[circle,draw] (3) at (0,-3) {4};
  \node[circle,draw] (4) at (3,-3) {3};
  \draw[blue, very thick] (1) -- (2);
  \draw[blue, very thick] (3) -- (4);
  \draw[red, very thick] (1) -- (3);
  \draw[red, very thick] (4) -- (2);
  \node (label0) at (-1.5,-2.5) {0};
  \node (label1) at (-1.5,-3) {1};
  \draw[blue,very thick] (label0)--(-1,-2.5);
  \draw[red,very thick] (label1)--(-1,-3);
    \end{tikzpicture}
    \caption{{A simple graph that corresponds to a source of postselected GHZ states \cite{Krenn_PhysRevLett.119.240403}.}}
    \label{fig:GHZ_graph_app}
\end{figure}
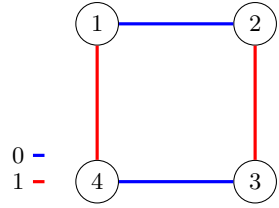
and the corresponding adjacency matrix, which has four unique nonzero elements, namely $w_{10,20}$, $w_{30,40}$, $w_{21,31}$, and $w_{41,11}$. For this state, the generation expansion given in Eq. \eqref{eq:state_big_main} is
\begin{align}
    \ket{\psi_2} &= \nonumber \bigg( 1 + 2\beta_{10;20} a^{\dagger}_{10} a^{\dagger}_{20} + 2\beta_{21;31} a^{\dagger}_{21} a^{\dagger}_{31} +  2\beta_{30;40} a^{\dagger}_{30} a^{\dagger}_{40} + 2\beta_{41;11} a^{\dagger}_{41} a^{\dagger}_{11}\\ &+ \nonumber 2(\beta_{10;20})^2 (a^{\dagger}_{10})^2 (a^{\dagger}_{20})^2 + 2(\beta_{21;31})^2 (a^{\dagger}_{21})^2 (a^{\dagger}_{31})^2 +  2(\beta_{30;40})^2 (a^{\dagger}_{30})^2 (a^{\dagger}_{40})^2 + 2(\beta_{41;11})^2 (a^{\dagger}_{41})^2 (a^{\dagger}_{11})^2
    \\ &+ \nonumber 4\beta_{10,20}\beta_{21,31} a^{\dagger}_{10} a^{\dagger}_{20} a^{\dagger}_{21} a^{\dagger}_{31}+ 4\beta_{10,20}\beta_{41,11} a^{\dagger}_{10} a^{\dagger}_{20} a^{\dagger}_{41} a^{\dagger}_{11} + 4\beta_{21,31}\beta_{30,40} a^{\dagger}_{21} a^{\dagger}_{31} a^{\dagger}_{30} a^{\dagger}_{40}\\ & + 4\beta_{30,40}\beta_{41,11}a^{\dagger}_{30} a^{\dagger}_{40} a^{\dagger}_{41} a^{\dagger}_{11} + 4\beta_{21,31}\beta_{41,11}a^{\dagger}_{21} a^{\dagger}_{31} a^{\dagger}_{41} a^{\dagger}_{11} + 4\beta_{10,20}\beta_{30,40}a^{\dagger}_{10} a^{\dagger}_{20} a^{\dagger}_{30} a^{\dagger}_{40}
    + ...\bigg) \ket{\text{vac}}. \label{eq:GHZ_full_example}
\end{align}
The sets of edges associated with each term are represented in Fig. \ref{fig:GHZ_decomp}. 
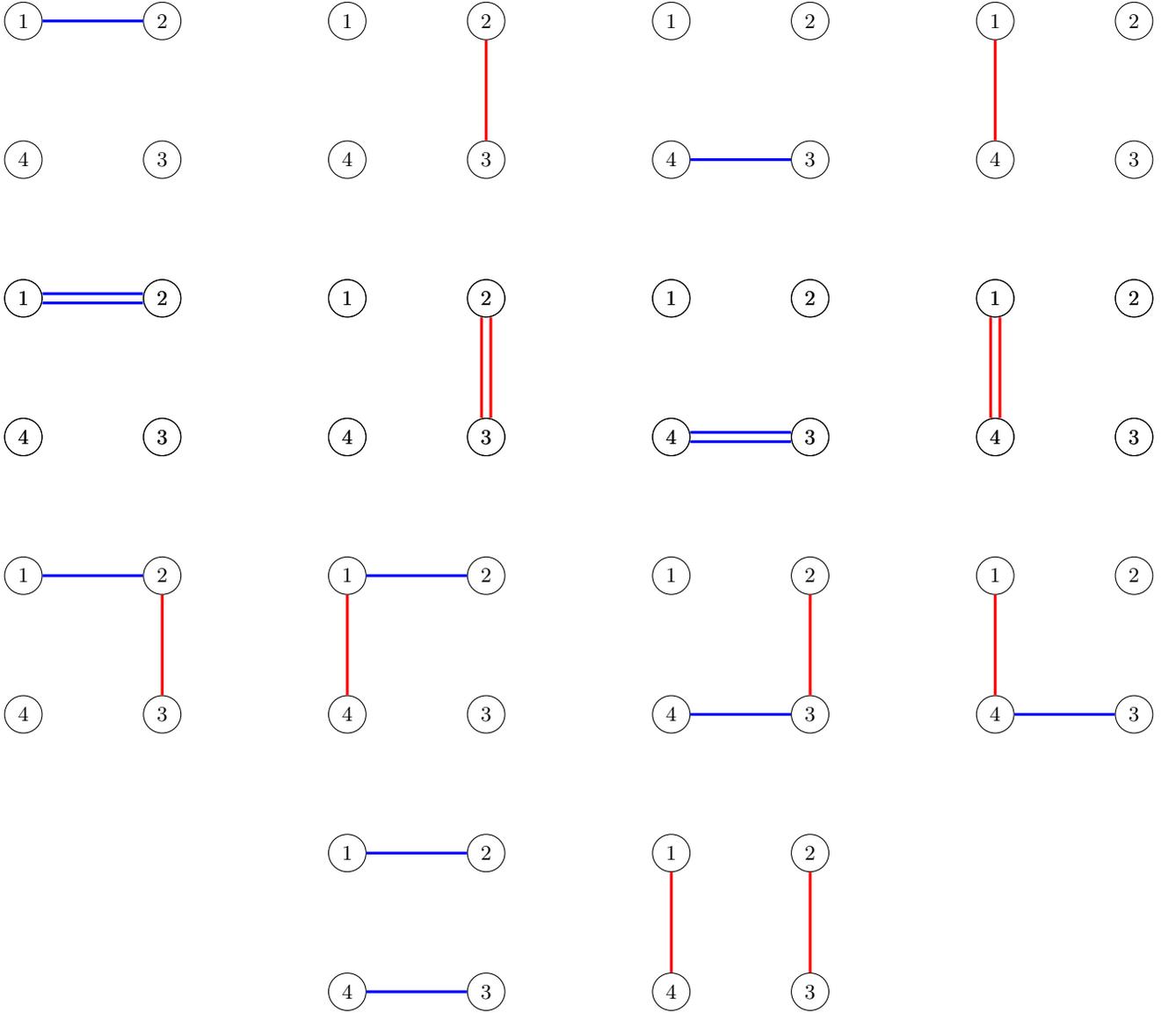
\begin{figure}[h]
    \centering
    \begin{tikzpicture}[scale=0.7]
  \node[circle,draw] (1) at (0,0) {1};
  \node[circle,draw] (2) at (3,0) {2};
  \node[circle,draw] (3) at (0,-3) {4};
  \node[circle,draw] (4) at (3,-3) {3};
  \node (5) at (1,-1) { };
  %
%
  \node[circle,draw] (1_2) at (7,0) {1};
  \node[circle,draw] (2_2) at (10,0) {2};
  \node[circle,draw] (3_2) at (7,-3) {4};
  \node[circle,draw] (4_2) at (10,-3) {3};
  \node[circle,draw] (1_3) at (14,0) {1};
  \node[circle,draw] (2_3) at (17,0) {2};
  \node[circle,draw] (3_3) at (14,-3) {4};
  \node[circle,draw] (4_3) at (17,-3) {3};
  \node[circle,draw] (1_4) at (21,0) {1};
  \node[circle,draw] (2_4) at (24,0) {2};
  \node[circle,draw] (3_4) at (21,-3) {4};
  \node[circle,draw] (4_4) at (24,-3) {3};
\node[circle,draw] (1_5) at (0,-6) {1};
  \node[circle,draw] (2_5) at (3,-6) {2};
  \node[circle,draw] (3_5) at (0,-9) {4};
  \node[circle,draw] (4_5) at (3,-9) {3};
  \node[circle,draw] (1_6) at (7,-6) {1};
  \node[circle,draw] (2_6) at (10,-6) {2};
  \node[circle,draw] (3_6) at (7,-9) {4};
  \node[circle,draw] (4_6) at (10,-9) {3};
  \node[circle,draw] (1_7) at (14,-6) {1};
  \node[circle,draw] (2_7) at (17,-6) {2};
  \node[circle,draw] (3_7) at (14,-9) {4};
  \node[circle,draw] (4_7) at (17,-9) {3};
  \node[circle,draw] (1_8) at (21,-6) {1};
  \node[circle,draw] (2_8) at (24,-6) {2};
  \node[circle,draw] (3_8) at (21,-9) {4};
  \node[circle,draw] (4_8) at (24,-9) {3};
\node[circle,draw] (1_5) at (0,-6) {1};
  \node[circle,draw] (2_5) at (3,-6) {2};
  \node[circle,draw] (3_5) at (0,-9) {4};
  \node[circle,draw] (4_5) at (3,-9) {3};
  \node[circle,draw] (1_6) at (7,-6) {1};
  \node[circle,draw] (2_6) at (10,-6) {2};
  \node[circle,draw] (3_6) at (7,-9) {4};
  \node[circle,draw] (4_6) at (10,-9) {3};
  \node[circle,draw] (1_7) at (14,-6) {1};
  \node[circle,draw] (2_7) at (17,-6) {2};
  \node[circle,draw] (3_7) at (14,-9) {4};
  \node[circle,draw] (4_7) at (17,-9) {3};
  \node[circle,draw] (1_8) at (21,-6) {1};
  \node[circle,draw] (2_8) at (24,-6) {2};
  \node[circle,draw] (3_8) at (21,-9) {4};
  \node[circle,draw] (4_8) at (24,-9) {3};
\node[circle,draw] (1_9) at (0,-12) {1};
  \node[circle,draw] (2_9) at (3,-12) {2};
  \node[circle,draw] (3_9) at (0,-15) {4};
  \node[circle,draw] (4_9) at (3,-15) {3};
  \node[circle,draw] (1_10) at (7,-12) {1};
  \node[circle,draw] (2_10) at (10,-12) {2};
  \node[circle,draw] (3_10) at (7,-15) {4};
  \node[circle,draw] (4_10) at (10,-15) {3};
  \node[circle,draw] (1_11) at (14,-12) {1};
  \node[circle,draw] (2_11) at (17,-12) {2};
  \node[circle,draw] (3_11) at (14,-15) {4};
  \node[circle,draw] (4_11) at (17,-15) {3};
  \node[circle,draw] (1_12) at (21,-12) {1};
  \node[circle,draw] (2_12) at (24,-12) {2};
  \node[circle,draw] (3_12) at (21,-15) {4};
  \node[circle,draw] (4_12) at (24,-15) {3};
  \node[circle,draw] (1_13) at (7,-18) {1};
  \node[circle,draw] (2_13) at (10,-18) {2};
  \node[circle,draw] (3_13) at (7,-21) {4};
  \node[circle,draw] (4_13) at (10,-21) {3};
  \node[circle,draw] (1_14) at (14,-18) {1};
  \node[circle,draw] (2_14) at (17,-18) {2};
  \node[circle,draw] (3_14) at (14,-21) {4};
  \node[circle,draw] (4_14) at (17,-21) {3};
  %
%
  \draw[blue, very thick] (1) -- (2);
  \draw[red,very thick] (2_2)--(4_2);
  \draw[blue, very thick] (3_3) -- (4_3); 
  \draw[red, very thick] (1_4) -- (3_4); 
    \draw[blue, very thick,transform canvas={yshift=2pt}] (1_5) -- (2_5);
    \draw[blue, very thick,transform canvas={yshift=-2pt}] (1_5) -- (2_5);
    \draw[red, very thick,transform canvas={xshift=2pt}] (2_6) -- (4_6);
    \draw[red, very thick,transform canvas={xshift=-2pt}] (2_6) -- (4_6);
    \draw[blue, very thick,transform canvas={yshift=2pt}] (3_7) -- (4_7);
    \draw[blue, very thick,transform canvas={yshift=-2pt}] (3_7) -- (4_7);
    \draw[red, very thick,transform canvas={xshift=2pt}] (1_8) -- (3_8);
    \draw[red, very thick,transform canvas={xshift=-2pt}] (1_8) -- (3_8);
    \draw[blue, very thick] (1_9) -- (2_9);
    \draw[red, very thick] (2_9) -- (4_9);
    \draw[blue, very thick] (1_10) -- (2_10);
    \draw[red, very thick] (1_10) -- (3_10);
    \draw[blue, very thick] (4_12) -- (3_12);
    \draw[red, very thick] (1_12) -- (3_12);
    \draw[blue, very thick] (4_11) -- (3_11);
    \draw[red, very thick] (2_11) -- (4_11);
    \draw[blue, very thick] (1_13) -- (2_13);
    \draw[blue, very thick] (4_13) -- (3_13);
    \draw[red, very thick] (1_14) -- (3_14);
    \draw[red, very thick] (2_14) -- (4_14);
\end{tikzpicture}
    \caption{The sets of edges that correspond to each term in Eq. \eqref{eq:GHZ_full_example}. The terms are represented in the order in which they are written in Eq. \eqref{eq:GHZ_full_example}.}
    \label{fig:GHZ_decomp}
\end{figure}

The measurement and postselection applied to the state can be understood as picking out a subset of the terms in Eq. \eqref{eq:GHZ_full_example}. Postselecting on at least one photon being detected in each of the output modes (i.e. four-fold coincidences) eliminates all the terms in Eq. \eqref{eq:GHZ_full_example} except for the last two; the other terms describe distributions of photons that leave at least one of the four output modes unoccupied. The postselected terms are associated with the sets of edges such that each vertex in the graph is connected to exactly one edge; by definition, these are the perfect matchings of the graph. 

To lowest order, the terms in the postselected state correspond to the perfect matchings of the graph that represents the multimode squeezed state prior to postselection. The postselected state also contains higher-order terms, but we take these to be negligible in the low-gain regime to which we restrict our discussion.


\pagebreak

\section{Source of frequency-bin-encoded $L_{a4}$ states}
 \label{appendix:L}

Here we apply the approach described in Sections \ref{section:notation} and \ref{section:graphs} to design a source of qubit $L_{a4}$ states, a class of entangled four-qubit states inequivalent to the four-photon GHZ and W states \cite{Verstraete_PhysRevA.65.052112}. We choose this state as a second example, partly to illustrate the ease with which two-coloured edges can be implemented in an integrated frequency-bin platform. 

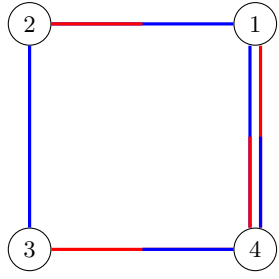
\begin{figure}[h!]  
\centering  
{
\begin{tikzpicture}
    \node (empty) at (-4,0){};
    \node (empty2) at (1,0){};
    \node[circle,draw] (2) at (-3,0) {2};
    \node[circle,draw] (1) at (0,0) {1};
    \node[circle,draw] (3) at (-3,-3) {3};
    \node[circle,draw] (4) at (0,-3) {4};
    \node (midpoint) at (0,-1.4) {};
    \node (center) at (-1.4,-1.4) {};
    \draw[red, very thick, transform canvas={xshift=2pt}] (1) -- (4) coordinate[pos=0.5] (midpoint3);
      \draw[blue, very thick, transform canvas={xshift=-2pt}] (1) -- (4);
      \draw[blue, very thick, transform canvas={xshift=2pt}] (midpoint3) -- (4);
     \draw[red, very thick, transform canvas={xshift=-2pt}] (midpoint3) -- (4);
      \draw[blue, very thick, transform canvas={xshift=0pt, yshift = 0pt}] (1) -- (2) coordinate [pos = 0.5] (midpoint1);
     \draw[red, very thick, transform canvas={xshift=0pt, yshift = 0pt}] (midpoint1) -- (2);
      \draw[blue, very thick, transform canvas={xshift=0pt}] (2) -- (3);
      \draw[red, very thick, transform canvas={xshift=0pt, yshift = 0pt}] (3) -- (4) coordinate [pos = 0.5] (midpoint5);
      \draw[blue, very thick, transform canvas={xshift=0pt, yshift = 0pt}] (midpoint5) -- (4);
\end{tikzpicture}}
\caption{Graph with perfect matchings that correspond to a postselected $L_{a4}$ state}
\label{fig:L_graph}
\end{figure} 

We take as our target the state
\begin{align}
    \ket{L_{a4}} = \alpha_1 \ket{0001} + \alpha_2 \ket{0110} + \alpha_3 \ket{1000}, \label{eq:L_ket}
\end{align}
where normalization requires that $|\alpha_1|^2 + |\alpha_1|^2 + |\alpha_1|^2 = 1 $. A graph with perfect matchings corresponding to this postselected state is given in Fig. \ref{fig:L_graph}, and the adjacency matrix corresponding to this graph can be written as
\begin{align}
    \boldsymbol{w} = \begin{bmatrix}
        \boldsymbol{w}_{00} & \boldsymbol{w}_{01} \\
        \boldsymbol{w}_{10} & \boldsymbol{w}_{11}
    \end{bmatrix},
\end{align}
with 
\begin{align}
    \boldsymbol{w}_{00} &= \begin{bmatrix}
        0 & 0 & 0 & 0 \\
        0 & 0 & \omega_{20;30} & 0 \\
        0 & \omega_{20;30} & 0 & 0 \\
        0 & 0 & 0 & 0
    \end{bmatrix},\\
    \boldsymbol{w}_{01} &= \begin{bmatrix}
        0 & \omega_{10;21} & 0 & \omega_{10;41} \\
        0 & 0 & 0 & 0 \\
        0 & 0 & 0 & 0 \\
        \omega_{40;11} & 0 & \omega_{40;31} & 0
    \end{bmatrix},\\
    \boldsymbol{w}_{10} &= \boldsymbol{w}^T_{01}, \\
    \boldsymbol{w}_{11} &= O_{4\times4},
\end{align}
where we use the notation of Section \ref{section:graphs}. The coefficients in Eq. \eqref{eq:L_ket} are given by 
\begin{align}
\alpha_1 = 4 w_{10;41} w_{20;30},\\
\alpha_2 = 4 w_{10;21} w_{40;31},\\
\alpha_3 = 4 w_{40;11} w_{20;30},
\end{align}
where the factors of 4 appear due to the symmetry of $\boldsymbol{w}$.

We put \begin{align}
    1 &\rightarrow \mathsf{a}{S},\\
    2 &\rightarrow \mathsf{b}{I},\\
    3 &\rightarrow \mathsf{b}{S},\\
    4 &\rightarrow \mathsf{a}{I},
\end{align}
and partition $\beta$ as 
\begin{align}
    \boldsymbol{\beta} = 
    \begin{bmatrix}
        O_{2\times2} & O_{2\times2} & \boldsymbol{\beta}_{S0,I0} & \boldsymbol{\beta}_{S0,I1} \\
        O_{2\times2} & O_{2\times2} & \boldsymbol{\beta}_{S1,I0} & O_{2\times2} \\
        \boldsymbol{\beta}_{I0,S0} & \boldsymbol{\beta}_{I0,S1} & O_{2\times2} & O_{2\times2} \\
        \boldsymbol{\beta}_{I1,S0} & O_{2\times2} & O_{2\times2} & O_{2\times2} \\
    \end{bmatrix}
\end{align}
\begin{align}
    \boldsymbol{\beta}_{S0,I0} &= \begin{bmatrix}
        0 & 0 \\
        0 & {\beta}_{\mathsf{b}S0;\mathsf{b}I0} 
    \end{bmatrix}, \\
    \boldsymbol{\beta}_{S0,I1} &= \begin{bmatrix}
        {\beta}_{\mathsf{a}S0;\mathsf{a}I1}  & {\beta}_{\mathsf{a}S0;\mathsf{b}I1} \\
        0 & 0
    \end{bmatrix}, \label{eq:L_S0I1}\\ 
    \boldsymbol{\beta}_{S1,I0} &= \begin{bmatrix}
        {\beta}_{\mathsf{a}S1;\mathsf{a}I0} & 0 \\
        {\beta}_{\mathsf{b}S1;\mathsf{a}I0} & 0 
    \end{bmatrix}, \label{eq:L_S1I0}\\
    \boldsymbol{\beta}_{n,n'} &= \left(\boldsymbol{\beta}_{n',n}\right)^T.
\end{align}

We seek solutions to 
\begin{align}
    \boldsymbol{\beta}_{ij} = \boldsymbol{U}_{ii} \boldsymbol{\overline{\beta}}_{ij} \boldsymbol{U}_{jj}^T,
\end{align}
for the nonzero blocks of $\boldsymbol{\beta}_{ij}$. Since $\boldsymbol{\beta}_{S0,I0}$ is diagonal, we can take 
\begin{align}
    \boldsymbol{U}_{S0,S0} &= \boldsymbol{U}_{I0,I0} = \mathbb{1}_{2\times2},\\
    \overline{\boldsymbol{\beta}}_{S0,I0} &= \begin{bmatrix}
        0 & 0 \\
        0 & {\beta}_{\mathsf{b}S0;\mathsf{b}I0} 
    \end{bmatrix}, \label{eq:beta_S0I0_L}
\end{align}
and so 
\begin{align}
    {\boldsymbol{\beta}}_{S0,I1} &= \overline{\boldsymbol{\beta}}_{S0,I1} \boldsymbol{U}_{I1,I1}^T \label{eq:L_1}\\
    {\boldsymbol{\beta}}_{S1,I0} &= \boldsymbol{U}_{S1,S1} \overline{\boldsymbol{\beta}}_{S1,I0} \label{eq:L_2}
\end{align}

At this point one can set the relative amplitudes of the elements of $\boldsymbol{\beta}$, and solve Eqs. \eqref{eq:L_1} and \eqref{eq:L_2} numerically with a polar decomposition. On the other hand, one can seek an analytic solution to determine the most general form required for the circuit. We place a further restriction on $\boldsymbol{\overline{\beta}}$: we seek a solution where the $2\times 2$ blocks $\boldsymbol{\overline{\beta}}_{nn'}$ are all diagonal, meaning the photon pair sources need to be coupled to only one output waveguide. We have 
\begin{align}
    {\boldsymbol{\beta}}_{S0,I1} &= \begin{bmatrix}
    \overline{\beta}_{\mathsf{a}S0,\mathsf{a}I1} & 0 \\ 0 & \overline{\beta}_{\mathsf{b}S0,\mathsf{b}I1}
    \end{bmatrix} \begin{bmatrix}
     {U}_{\mathsf{a}I1,\mathsf{a}I1} & {U}_{\mathsf{b}I1,\mathsf{a}I1} \\ {U}_{\mathsf{a}I1,\mathsf{b}I1} & {U}_{\mathsf{b}I1,\mathsf{b}I1} 
    \end{bmatrix}
    = \begin{bmatrix}
        {U}_{\mathsf{a}I1,\mathsf{a}I1} \overline{\beta}_{\mathsf{a}S0,\mathsf{a}I1} & {U}_{\mathsf{b}I1,\mathsf{a}I1}\overline{\beta}_{\mathsf{a}S0,\mathsf{a}I1} \\ {U}_{\mathsf{a}I1,\mathsf{b}I1}\overline{\beta}_{\mathsf{b}S0,\mathsf{b}I1} & {U}_{\mathsf{b}I1,\mathsf{b}I1}\overline{\beta}_{\mathsf{b}S0,\mathsf{b}I1}
    \end{bmatrix}, \label{eq:beta_S0I1_L}
\end{align}
and 
\begin{align}
    {\boldsymbol{\beta}}_{S1,I0} &= \begin{bmatrix}
     {U}_{\mathsf{a}S1,\mathsf{a}S1} & {U}_{\mathsf{a}S1,\mathsf{b}S1} \\ {U}_{\mathsf{b}S1,\mathsf{a}S1} & {U}_{\mathsf{b}S1,\mathsf{b}S1} 
    \end{bmatrix} \begin{bmatrix}
    \overline{\beta}_{\mathsf{a}S1,\mathsf{a}I0} & 0 \\ 0 & \overline{\beta}_{\mathsf{b}S1,\mathsf{b}I0}
    \end{bmatrix}
    = \begin{bmatrix}
        {U}_{\mathsf{a}S1,\mathsf{a}S1} \overline{\beta}_{\mathsf{a}S1,\mathsf{a}I0} & {U}_{\mathsf{a}S1,\mathsf{b}S1}\overline{\beta}_{\mathsf{b}S1,\mathsf{b}I0} \\ {U}_{\mathsf{b}S1,\mathsf{a}S1}\overline{\beta}_{\mathsf{a}S1,\mathsf{a}I0} & {U}_{\mathsf{b}S1,\mathsf{b}S1}\overline{\beta}_{\mathsf{b}S1,\mathsf{b}I0}
    \end{bmatrix}. \label{eq:beta_S1I0_L}
\end{align}
Comparing with Eqs. \eqref{eq:L_S0I1} and \eqref{eq:L_S1I0}, we have 
\begin{align}
    \beta_{\mathsf{a}S0,\mathsf{a}I1} &= {U}_{\mathsf{a}I1,\mathsf{a}I1} \overline{\beta}_{\mathsf{a}S0,\mathsf{a}I1} \label{eq:beta_S0I1_L_2}\\
    \beta_{\mathsf{a}S0,\mathsf{b}I1} &= {U}_{\mathsf{b}I1,\mathsf{a}I1}\overline{\beta}_{\mathsf{a}S0,\mathsf{a}I1}\\
    \beta_{\mathsf{a}S1,\mathsf{a}I0} &= {U}_{\mathsf{a}S1,\mathsf{a}S1} \overline{\beta}_{\mathsf{a}S1,\mathsf{a}I0}\\
    \beta_{\mathsf{b}S1,\mathsf{a}I0} &= {U}_{\mathsf{b}S1,\mathsf{a}S1}\overline{\beta}_{\mathsf{a}S1,\mathsf{a}I0}, \label{eq:beta_S1I0_L_2}
\end{align}
and 
\begin{align}
    \overline{\beta}_{\mathsf{b}S0,\mathsf{b}I1} &= \overline{\beta}_{\mathsf{b}S1,\mathsf{b}I0} = 0.
\end{align}

With this we have the general form that is required for the sources and linear components; the device is sketched in Fig. \ref{fig:L_device}.
\begin{figure}[h]
    \centering
    \subfloat[]{
        \includegraphics[width=0.6\textwidth]{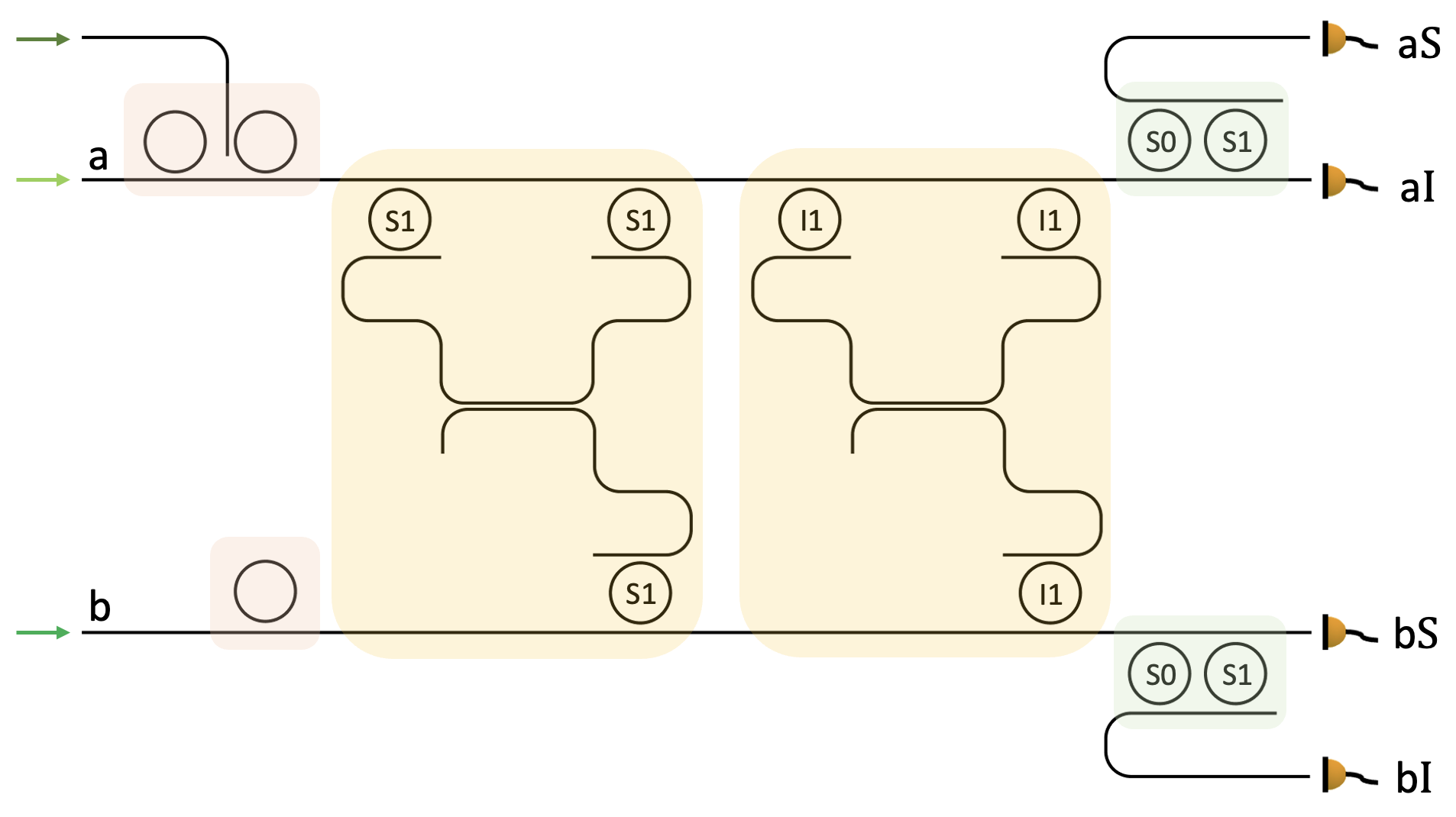}
    }

    \subfloat[]{
        \includegraphics[width=0.6\textwidth]{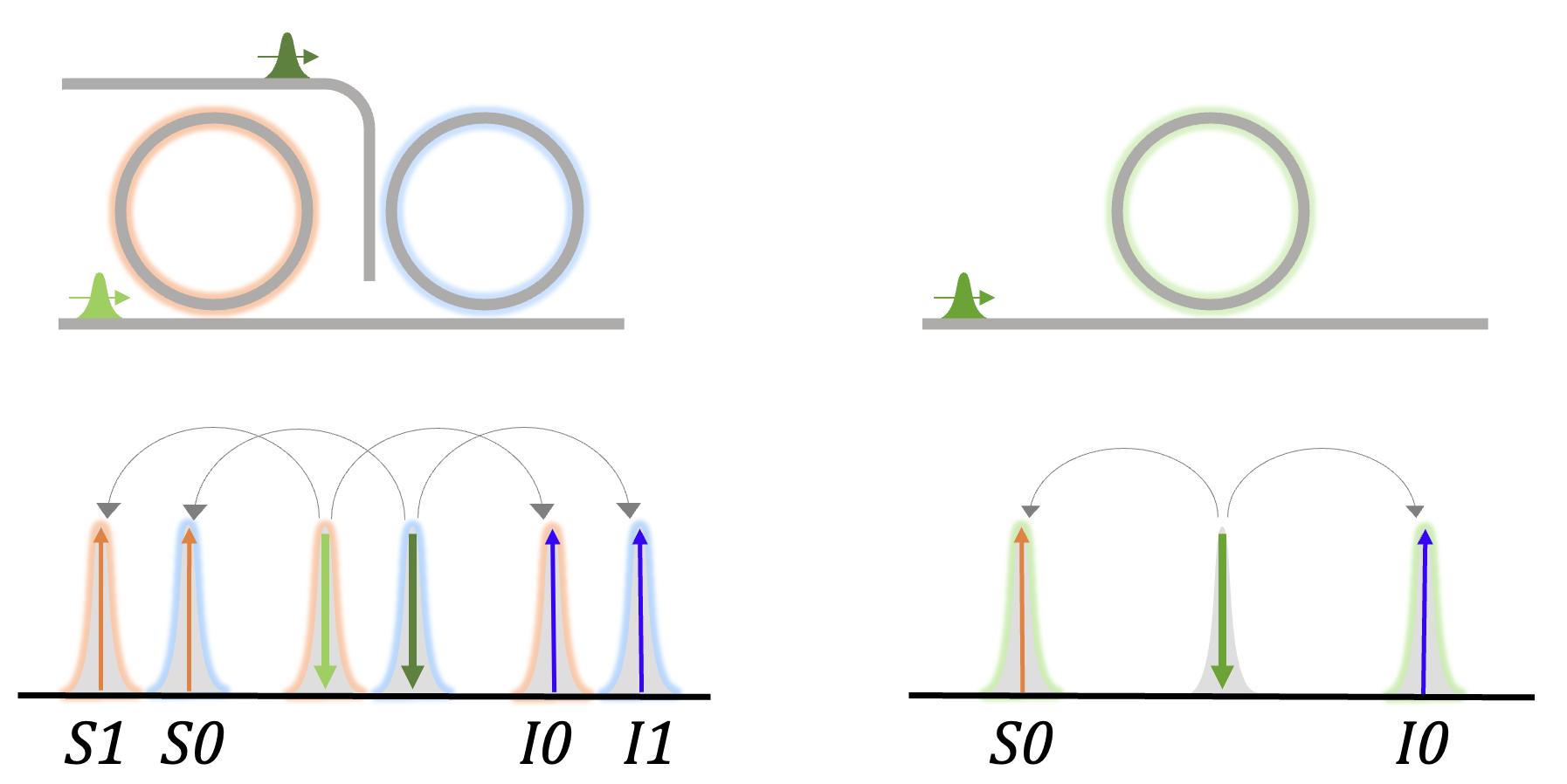}
    }
    \caption{Sketch of a passive integrated source of postselected L states (a). The arrows indicate the input pump fields, and the detectors indicate output ports. The sources are highlighted in orange; the configurations of these are indicated in panel (b). The linear components are highlighted in yellow and green.}
    \label{fig:L_device}
\end{figure}

The photon pair sources indicated in Fig. \ref{fig:L_device} are configured as described by the diagonal blocks of $\boldsymbol{\overline{\beta}}$ appearing in Eqs. \eqref{eq:beta_S0I0_L}, \eqref{eq:beta_S0I1_L}, and \eqref{eq:beta_S1I0_L}. The sections highlighted in yellow act as frequency-dependent DCs; only photons at the microrings' resonant frequencies (labelled on the rings in Fig. \ref{fig:L_device}a) propagate through the DC. The reflection and transmission coefficients depend on the values of the amplitudes in Eqs. \eqref{eq:beta_S0I1_L_2} to \eqref{eq:beta_S1I0_L_2}. For example, the DCs should act as 50-50 beamsplitters to generate an $L$ state with $|\alpha_1| = |\alpha_2| = |\alpha_3|$ in Eq. \eqref{eq:L_ket}. A numerical polar decomposition yields the same results. 


\end{document}